\documentclass[sigconf]{acmart}

\AtBeginDocument{%
  }

\copyrightyear{2026}
\acmYear{2026}
\setcopyright{cc}
\setcctype{by}
\acmConference[HPDC '26]{The 35th International Symposium on High-Performance Parallel and Distributed Computing}{July 13--16, 2026}{Cleveland, OH, USA}
\acmBooktitle{The 35th International Symposium on High-Performance Parallel and Distributed Computing (HPDC '26), July 13--16, 2026, Cleveland, OH, USA}
\acmDOI{10.1145/3806645.3807584}
\acmISBN{979-8-4007-2640-8/2026/07}

\usepackage{graphicx}  
\usepackage{algorithm}
\usepackage{algpseudocode}
\usepackage{enumitem}
\usepackage{xcolor}
\usepackage{pifont}
\usepackage{subcaption}
\usepackage{multirow}

\begin{document}

\title[TACO]{TACO: Efficient Communication Compression of Intermediate Tensors for Scalable Tensor-Parallel LLM Training}
%Efficient Communication Compression of Intermediate Tensors for Scalable Tensor-Parallel LLM Training

\author{Man Liu}
\affiliation{%
\institution{Hangzhou Institute for Advanced Study, University of Chinese Academy of Sciences}
\city{Hangzhou}
\country{China}}
\email{liuman24@mails.ucas.ac.cn}

\author{Xingchen Liu}
\affiliation{%
\institution{Institute of Computing Technology, Chinese Academy of Sciences}
\city{Beijing}
\country{China}
}
\email{liuxingchen23s@ict.ac.cn}

\author{Xingjian Tian}
\affiliation{%
\institution{Hangzhou Institute for Advanced Study, University of Chinese Academy of Sciences}
\city{Hangzhou}
\country{China}}
\email{tianxingjian25@mails.ucas.ac.cn}

\author{Bing Lu}
\affiliation{%
\institution{Institute of Computing Technology, Chinese Academy of Sciences}
\city{Beijing}
\country{China}}
\email{lubing@ict.ac.cn}

\author{Shengkai Lyu}
\affiliation{%
\institution{Institute of Computing Technology, Chinese Academy of Sciences}
\city{Beijing}
\country{China}}
\email{lvshengkai23@mails.ucas.ac.cn}

\author{Shengquan Yin}
\affiliation{%
\institution{University of Science and Technology of China}
\city{Hefei}
\country{China}}
\email{yinshengquan@mail.ustc.edu.cn}

\author{Wenjing Huang}
\affiliation{%
\institution{Institute of Computing Technology, Chinese Academy of Sciences}
\city{Beijing}
\country{China}}
\email{huangwenjing23@mails.ucas.ac.cn}

\author{Zheng Wei}
\affiliation{%
\institution{Institute of Computing Technology, Chinese Academy of Sciences}
\city{Beijing}
\country{China}}
\email{weizheng@ncic.ac.cn}

\author{Hairui Zhao}
\affiliation{%
\institution{Institute of Computing Technology, Chinese Academy of Sciences}
\city{Beijing}
\country{China}}
\email{zhaohairui@ict.ac.cn}

\author{Guangming Tan}
\affiliation{%
\institution{Institute of Computing Technology, Chinese Academy of Sciences}
\city{Beijing}
\country{China}}
\email{tgm@ict.ac.cn}

\author{Dingwen Tao}
\affiliation{%
\institution{Institute of Computing Technology, Chinese Academy of Sciences}
\city{Beijing}
\country{China}}
\email{taodingwen@ict.ac.cn}

\renewcommand{\shortauthors}{Liu et al.}

%\renewcommand{\shortauthors}{Liu et al.}

\begin{abstract}
Handling communication overhead in large-scale tensor-parallel training remains a critical challenge due to the dense, near-zero distributions of intermediate tensors, which exacerbate errors under frequent communication and introduce significant computational overhead during compression. To this end, we propose \textbf{TACO} (\textbf{T}ensor-parallel \textbf{A}daptive \textbf{CO}mmunication compression), a robust FP8-based framework for compressing TP intermediate tensors. First, we employ a data-driven reshaping strategy combined with an Adaptive Scale–Hadamard Transform to enable high-fidelity FP8 quantization, while its Dual-Scale Quantization mechanism ensures numerical stability throughout training. Second, we design a highly fused compression operator to reduce memory traffic and kernel launch overhead, allowing efficient overlap with communication. Finally, we integrate TACO with existing state-of-the-art methods for Data and Pipeline Parallelism to develop a compression-enabled 3D-parallel training framework. Detailed experiments on GPT models and Qwen model demonstrate up to 1.87$\times$ end-to-end throughput improvement while maintaining near-lossless accuracy, validating the effectiveness and efficiency of TACO in large-scale training.
\end{abstract}




\maketitle

% ====================================================================================================
% Introduction
% ====================================================================================================

\section{Introduction}

The rapid scaling of large language models (LLMs) to tens of billions, hundreds of billions, and even trillion-parameter scales has driven the adoption of increasingly sophisticated distributed training strategies~\cite{lamprecht2025, PaLM, Jiang2024MegaScaleSL, Smith2022UsingDA}. Among them, \textit{3D parallelism} — comprising data parallelism (DP), tensor parallelism (TP), and pipeline parallelism (PP)—has emerged as the dominant paradigm for training ultra-large models~\cite{3DZheng2024, chen2024eellm}. However, as model scale grows, training performance becomes increasingly constrained by communication rather than computation. Recent system studies show that communication can account for over 50\% of the total training time~\cite{narayanan2021efficient, wang2024hiding}. Different parallelism strategies exhibit fundamentally different communication patterns and compression challenges. DP performs relatively low-frequency gradient synchronization, while PP primarily relies on lightweight point-to-point communication without global synchronization, making both comparatively amenable to effective communication compression~\cite{2020zero, alistarh2017qsgd, huang2019gpipe, narayanan2019pipedream}.

\begin{figure}[t]
    \centering
    \begin{subfigure}[t]{0.34\linewidth}
        \centering
        \includegraphics[width=\linewidth]{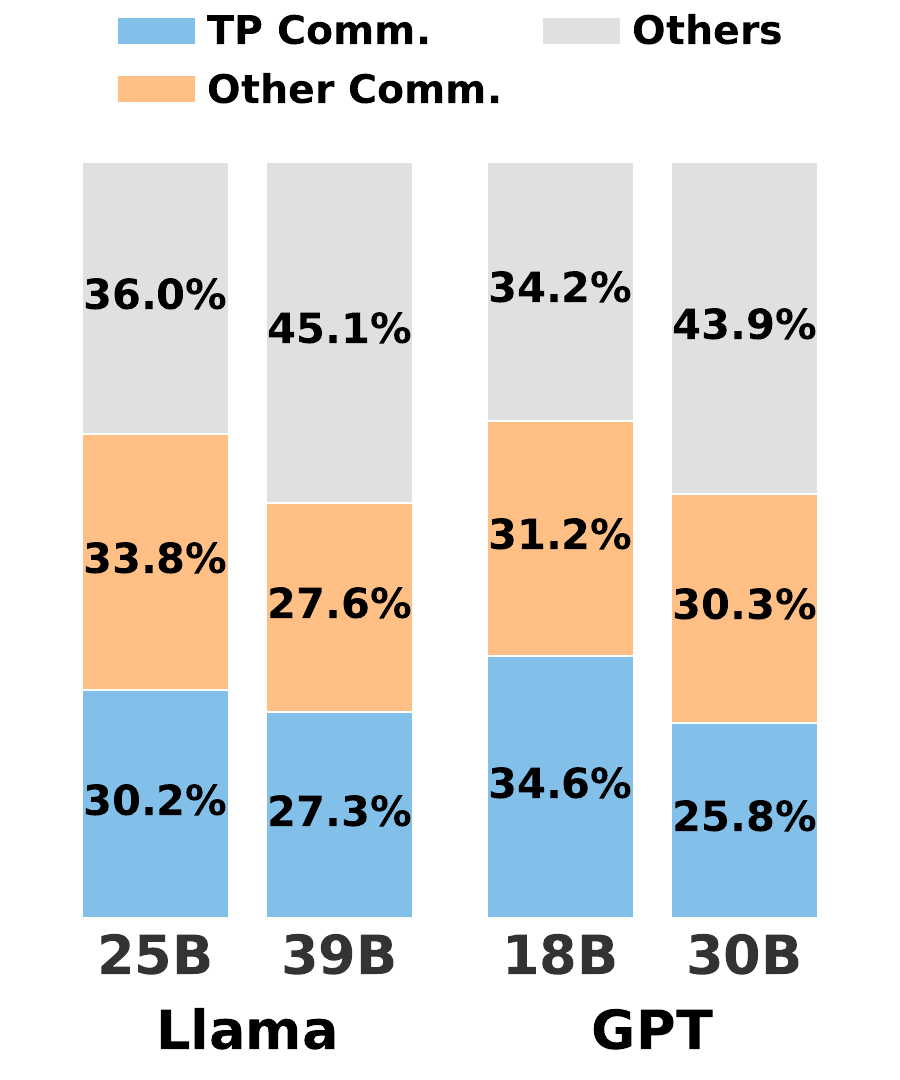}
        \caption{Execution time breakdown across different models and scales}
        \label{fig:tp_time}
    \end{subfigure}
    \hfill
    \begin{subfigure}[t]{0.65\linewidth}
        \centering
        \includegraphics[width=\linewidth]{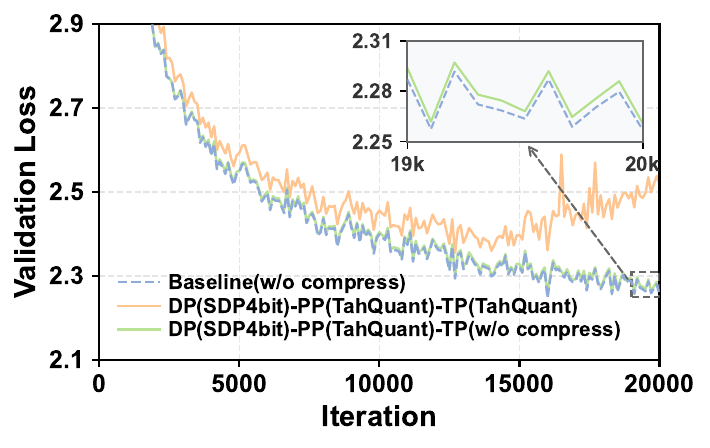}
        \caption{Validation loss comparison between the baseline and TP using TahQuant compression under 3D parallelism on GPT-350M}
        \label{fig:loss}
    \end{subfigure}
    \caption{Communication overhead and impact of quantization on training performance and convergence}
    \Description{The left plot shows execution time breakdown across different models and scales. The right plot compares validation loss between baseline and tensor parallel training with TahQuant compression under 3D parallelism.}

    \label{fig:tp_and_loss}
\end{figure}

In contrast, TP requires frequent, tightly synchronized communication to exchange intermediate tensors (sharded activations and gradients) during both forward and backward passes~\cite{shoeybi2019megatronlm, Brown2020gpt3, touvron2023llama, grattafiori2024llama3, Scao2022BLOOMA1}. As a result, TP communication lies on the critical execution path, accounts for over 50--60\% of total communication time (see Figure~\ref{fig:tp_time}), and is notoriously difficult to overlap with computation~\cite{narayanan2021efficient, shoeybi2019megatronlm, wang2024hiding}.This creates a fundamental challenge—\textbf{compression overhead under frequent communication}—as compression overhead must be minimized, which demands extreme operator-level optimization while carefully and strictly controlling error.

\sloppy
To mitigate communication overhead, prior work has explored various compression techniques, including quantization~\cite{zhang2023lowbit, qsdp2023}, sparsification~\cite{ScaleCom2020, li2022ppopp}, and low-rank approximation~\cite{PowerSGD, zhang2023lowbit}. These approaches have been successfully applied in DP and PP. Representative methods such as SDP4bit~\cite{jia2024sdp4bit} and TahQuant~\cite{he2025tahquant} demonstrate that carefully designed compression strategies can significantly reduce communication cost while effectively preserving training stability in their respective domains~\cite{QLORA2023}. In parallel, many inference-oriented compression methods—such as SmoothQuant~\cite{xiao2024smoothquant}, FlatQuant~\cite{sun2025flatquant}, QuaRot~\cite{ashkboos2024quarot}, and GPTQ~\cite{frantar2023gptq}—primarily leverage INT4/INT8 formats tailored for \textit{static weights and activations}.

However, communication compression for TP remains largely underexplored, particularly in training. Unlike the coarse-grained and relatively infrequent communication patterns in DP and PP, TP necessitates multiple rounds of compression within every Transformer block communication, giving rise to a critical challenge: \textbf{error accumulation under high-frequency communication}. Even minor quantization errors in TP intermediate tensors can propagate through attention and residual connections, amplifying during backpropagation~\cite{xu2024accelerating} and, as shown in Figure~\ref{fig:loss}, potentially destabilizing training or causing divergence. 
Moreover, TP communication compression is much more challenging during training than inference. Unlike inference, pretraining involves thousands of forward and backward passes, repeatedly injecting and amplifying quantization noise, so directly applying inference-oriented compression to TP tensors often leads to catastrophic training failure.

Through a systematic analysis of the distributions of TP intermediate tensors, we make two key observations that clearly and fundamentally highlight the fundamental difficulty of their compression. First, these tensors are strongly dominated by \textit{small-magnitude values with highly concentrated, zero-centered distributions}. This characteristic constitutes the primary challenge for TP communication compression (\textbf{distinct TP intermediate tensor characteristics}), as conventional methods fail to adequately capture the numerical subtleties of these extremely small values, resulting in significant information loss and progressively severe error accumulation, especially under repeated synchronization along the TP computation path.
Second, regarding quantization preprocessing, prior work commonly applies fixed transformations—such as rotations or Hadamard transforms—to balance value distributions~\cite{chee2023quip, tseng2024quip, ashkboos2024quik, savkin2025spinquant, egiazarian2025bridging}. While these transforms can spread variance across dimensions in other contexts, they are data-independent and lack the adaptability required to disperse the dense, zero-centered clusters inherent in TP intermediate tensors. Consequently, static transforms alone are insufficient to enable high-fidelity and reliable quantization for tightly synchronized TP communication.

\textit{Motivated by these challenges}, we propose \textbf{TACO} (\textbf{T}ensor-parallel \textbf{A}daptive \textbf{CO}mmunication compression), a robust FP8-based framework explicitly designed to efficiently compress intermediate tensors within TP training. 
\textit{TACO is specifically designed to address the critical challenges of error accumulation during TP pretraining, the distinct characteristics of TP intermediate tensors, and the high compression overhead induced by tightly synchronized communication.}
Unlike prior works, TACO directly targets latency-critical intermediate tensor communication in TP training, where both numerical fidelity and efficiency are essential.
It introduces a data-driven reshaping strategy that dynamically adapts to the statistical distributions of TP intermediate tensors, enabling high-fidelity FP8 quantization in practice. To ensure numerical stability, the framework employs a dual-scale quantization technique that preserves precision across all training stages. In addition, we develop a highly fused compression operator to minimize memory traffic and kernel launch overhead, significantly enhancing computational efficiency. To the best of our knowledge, TACO is \textbf{the first 3D parallel communication compression system} that achieves stable convergence in large-scale TP training scenarios. Our primary contributions are summarized as follows.

\begin{itemize}[leftmargin=1.3em,topsep=2pt]
    \item We systematically analyze TP intermediate tensors, whose \textbf{dense, near-zero distributions} render INT8 quantization inadequate while favoring FP8. Nevertheless, FP8’s representational capacity under low-bit remains limited, and standard Hadamard transform fails to sufficiently disperse these highly concentrated values. This analysis provides crucial guidance for designing effective TP intermediate tensor compression methods.

    \item We propose \textbf{TACO}, a TP communication compression framework. TACO introduces the \textit{Adaptive Scale--Hadamard Transform} to dynamically reshape the distribution of TP intermediate tensors for high-fidelity FP8 quantization, and implements \textit{Dual-Scale Quantization} to keep all communicated values within the FP8 representable range across training, effectively \textbf{preventing error amplification under frequent TP communication}.
    
    \item We develop \textbf{a highly fused compression operator} that combines all quantization steps in a single kernel, reuses warp-level reductions, and accesses metadata without extra copies. This design reduces memory traffic and kernel launches, accelerates computation, and enables efficient overlap with communication via tight integration with backends and fine-grained scheduling, thus \textbf{mitigating high-frequency compression overhead}.

    \item We integrate TACO with state-of-the-art compression methods for DP (SDP4Bit) and PP (TahQuant), enabling fully compression-enabled 3D-parallel training. In addition, evaluation results on GPT models and Qwen2.5-7B show that our method achieves up to \textbf{1.87×} end-to-end throughput improvement for TP compression and up to \textbf{1.53×} under full 3D-parallel compression, while \textbf{maintaining near-lossless training accuracy}.

\end{itemize}

\begin{figure}[t]
    \centering
    \includegraphics[width=1.0\linewidth]{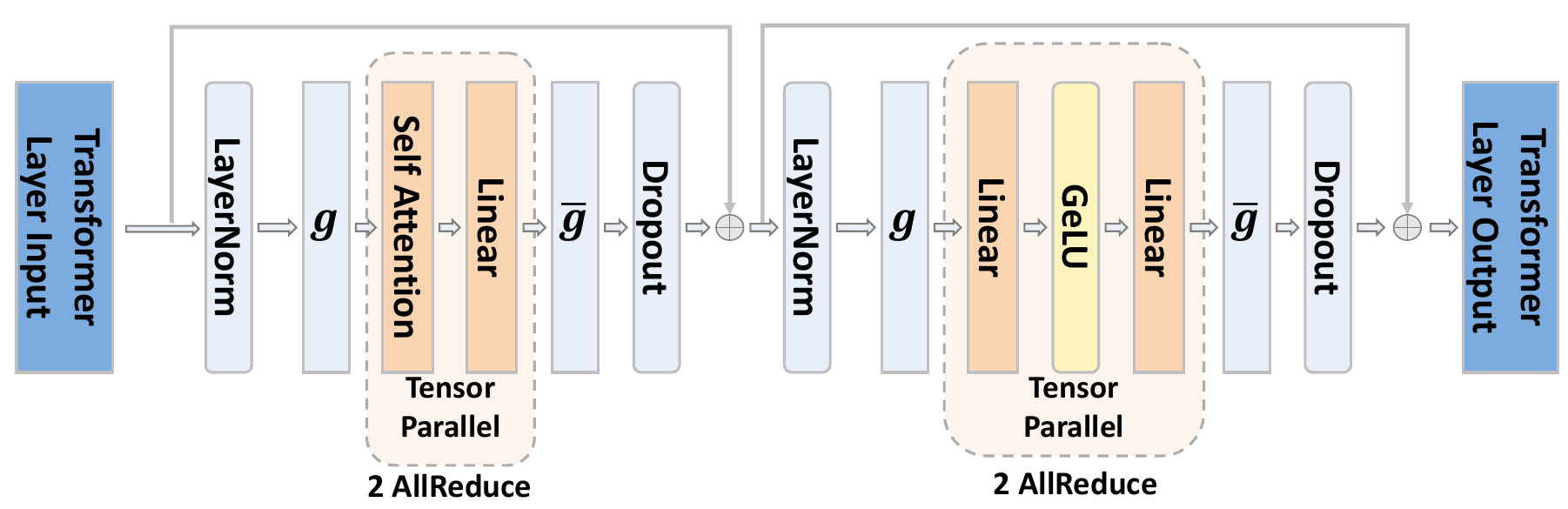}
    \caption{Illustration of TP in a Transformer block, including the AllReduce operations required during forward and backward passes.}
    \Description{Diagram of tensor parallelism in a Transformer block showing how computation is split across devices and where AllReduce operations occur during forward and backward passes.}
    \label{fig:tp_diagram}
\end{figure}

% ====================================================================================================
% Background and Related Work
% ====================================================================================================
\section{Background and Related Work}

\subsection{Tensor Parallelism and Its Communication}

Currently, distributed training of LLMs encounters increasingly significant communication bottlenecks stemming from different parallelism strategies, including DP, PP, and TP. Among these, TP is the primary contributor to intra-node communication overhead, as it requires frequent synchronization of intermediate activations and gradients across GPUs. 
TP enables the scaling of Transformer-based LLMs across multiple GPUs by exploiting the inherent parallelism of large matrix multiplications~\cite{shoeybi2019megatronlm, IEEE2024}. In TP, weight matrices are partitioned along the hidden dimension—either by rows or columns—allowing multiple devices to collaboratively compute a single Transformer block. As illustrated in Figure~\ref{fig:tp_diagram}, mainstream implementations apply TP to both the MLP and Attention blocks. During the forward pass, column-parallel linear layers produce partial outputs on each GPU, which must be synchronized via collective communication before being consumed by subsequent layers. Similarly, during backpropagation, row-parallel layers require collective synchronization of gradients prior to parameter updates.

Because TP collectives are invoked synchronously at every layer and iteration, TP constitutes the primary intra-node communication pathway in 3D-parallel training systems. Whether implemented as \texttt{AllReduce} or, when combined with sequence parallelism (SP), decomposed into \texttt{AllGather} and \texttt{Reduce-Scatter} operations along the sequence dimension~\cite{li2023sequenceparallelism}, TP communication is highly sensitive to both bandwidth and latency. As model widths scale to tens of thousands and depths grow to hundreds of layers, the communication volume incurred by TP increases rapidly, leading to severe bottlenecks even on advanced GPU interconnects~\cite{shoeybi2019megatronlm, narayanan2021efficient, Brown2020gpt3, gale2022megablocks}. Consequently, TP communication consistently accounts for a substantial fraction of the overall training communication cost.

\begin{figure}[t]
  \centering
  \includegraphics[width=0.9\linewidth]{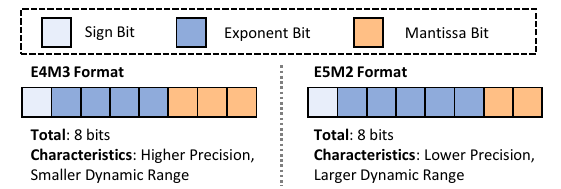}
  \caption{FP8 Format Specifications: E4M3 vs. E5M2. E4M3 provides a larger dynamic range with relatively lower precision, while E5M2 offers higher precision but a smaller dynamic range.}
  \Description{Comparison of two FP8 formats. E4M3 uses fewer exponent bits and more mantissa bits, providing higher precision but a smaller dynamic range. E5M2 uses more exponent bits and fewer mantissa bits, resulting in a larger dynamic range but lower precision.}
  \label{fig:e4m3/e5m2_format}
\end{figure}

\subsection{Communication Compression in Distributed Training}

Communication compression has been extensively studied in distributed training, particularly for data-parallel  gradient synchronization. In DP, gradients exhibit relatively stable distributions and tolerate moderate approximation errors, enabling diverse compression techniques, including stochastic quantization~\cite{alistarh2017qsgd}, low-bit gradient methods~\cite{lin2017deep}, and low-precision optimizers~\cite{zhang2023lowbit, qsdp2023, QLORA2023}. Quantization-aware training methods, such as LLM-QAT~\cite{llmqat2024}, further enhance robustness under low-precision representations by incorporating quantization effects during training, mainly targeting weights and activations. 
More recent system-level designs further push DP gradient communication toward ultra-low precision. For example, SDP4Bit achieves near-4-bit communication in sharded DP training~\cite{jia2024sdp4bit}, EDGC uses entropy-driven dynamic gradient compression to reduce latency in GPT training~\cite{yi2025edgc}, and TAGC introduces a Transformer-aware hierarchical compression scheme~\cite{polyakov2025tagc}. Collectively, these methods demonstrate that aggressive communication compression is feasible in DP while preserving convergence.

Beyond DP, communication compression has also been studied in PP training, where reducing inter-stage communication can significantly improve throughput. Representative approaches include dynamic precision control~\cite{chen2021actnn}, activation difference compression~\cite{wang2022AQSGD}, and joint activation–gradient compression with error compensation~\cite{rudakov2023activations}. More recently, TahQuant introduces fine-grained activation quantization along the PP communication path to improve accuracy preservation~\cite{he2025tahquant}. While effective in PP settings, these methods typically rely on relatively loose synchronization constraints.
Recent system-level efforts further explore communication-efficient training via coordinated compression and execution optimization, e.g., Tango~\cite{tango2023}, which reduces computation and communication overhead through quantization-aware scheduling.

In contrast, TP exhibits fundamentally different communication characteristics. TP intermediate tensors are smaller, exchanged at higher frequency, and reside on the critical computation path. Consequently, TP communication is highly sensitive to numerical perturbations, and directly applying DP- or PP-oriented compression methods can severely degrade numerical stability and hinder convergence. Although recent efforts explore low-bit TP communication~\cite{dong2024towards, li2024flash}, they primarily target inference workloads, which are more tolerant of approximation errors.
Furthermore, in end-to-end 3D-parallel training systems, existing designs often adopt conservative strategies that leave TP communication uncompressed to avoid catastrophic training divergence~\cite{xu2024accelerating}. 
To date, there remains no communication compression scheme that enables stable convergence for TP training of large language models.

\subsection{Numerical Challenges of Low-Bit Quantization in TP Communication}

Low-bit quantization is a common strategy for reducing communication and memory overhead in large-scale model training. In practice, INT8 quantization is widely adopted in inference due to its simplicity and favorable efficiency--accuracy trade-offs~\cite{sun2025flatquant, ashkboos2024quarot}. However, its applicability to TP communication during pretraining is constrained by much stricter numerical stability requirements. INT8 maps floating-point values to a uniform integer grid using a single scaling factor, with quantization defined as $Q_{\text{INT8}}(x)=\mathrm{round}(x/\Delta)$ and $\Delta=|x|_{\max}/127$. While hardware-efficient, this uniform quantization provides limited effective resolution for values concentrated near zero, which are frequently observed in TP intermediate tensors~\cite{kuzmin2022fp8quantization}.
In high-frequency, tightly synchronized TP communication, small mismatches in scaling factors across layers or shards can introduce rounding and saturation effects, leading to distorted gradient aggregation and unstable optimization dynamics~\cite{Micikevicius2018fp8}.

In contrast, reduced-precision floating-point formats, such as FP8, preserve scale information through explicit exponent encoding, enabling more robust handling of heterogeneous and dynamically changing tensor values. An FP8 number is represented as $x = (-1)^s \cdot 2^{e-B} \cdot (1+f)$, where $s$, $e$, and $f$ denote the sign, exponent, and mantissa, respectively.
Common FP8 variants include E4M3 and E5M2~\cite{shen2024efficient}. The bit-level structures of these variants are illustrated in Figure~\ref{fig:e4m3/e5m2_format}.This inherent adaptivity makes FP8 particularly well-suited for low-bit TP communication during pretraining, addressing key numerical limitations of INT8 in such settings~\cite{micikevicius2022fp8}.
Moreover, FP8 is currently being widely adopted in modern AI accelerators, including NVIDIA and AMD GPUs, and is expected to become a broadly supported type in the future—similar to how FP16 gained widespread hardware adoption—making it a timely and forward-looking choice for large-scale training.

While FP8 mitigates several numerical limitations inherent to INT8 quantization, effective TP communication compression depends not only on the numerical format itself, but also critically on how well it aligns with the statistical properties and synchronization patterns of TP intermediate tensors. In the following section, we analyze why FP8 is particularly well suited for TP communication and how its representational characteristics can be leveraged to achieve both high compression efficiency and stable training.

\begin{figure}[t]
    \centering
    \includegraphics[width=0.9\linewidth]{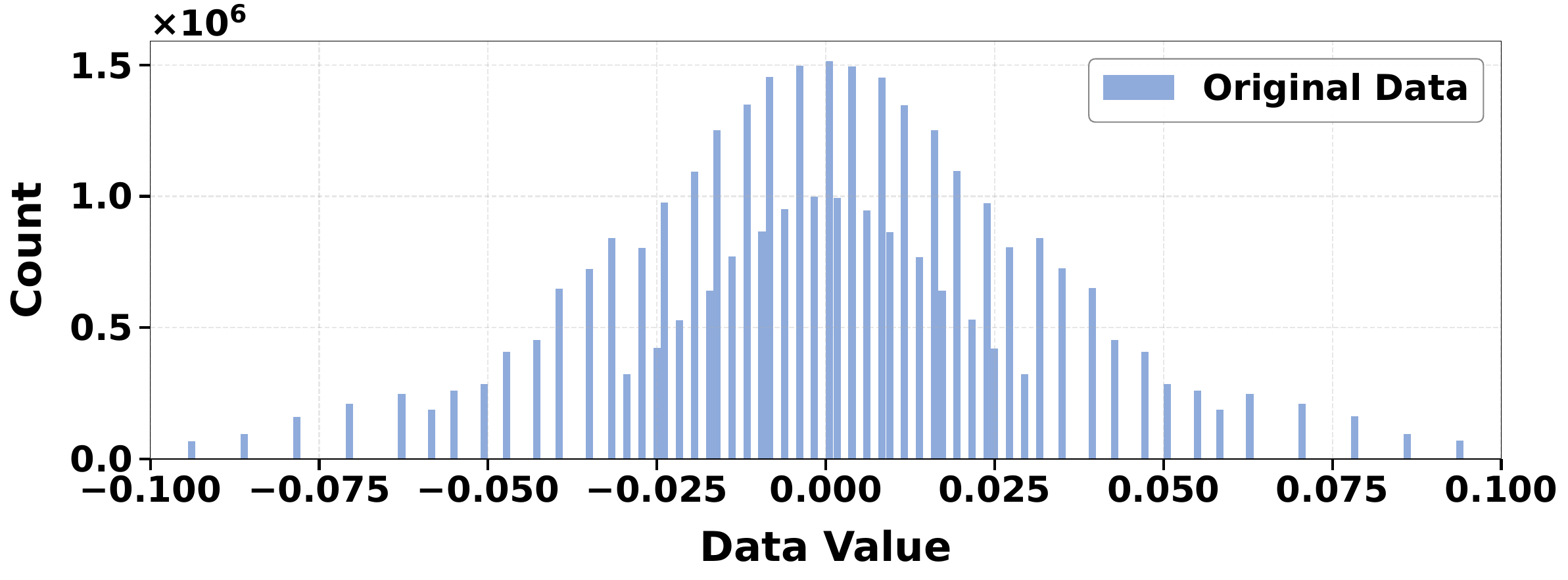}
    \caption{Histogram of TP communication data.}
    \Description{Histogram showing the distribution of tensor parallel communication data values, illustrating how frequently different value ranges occur and revealing a skewed distribution pattern.}
    \label{fig:original_hist}
\end{figure}

% ==================================================================
% Section 3: Analysis
% ==================================================================

\section{Why FP8 is Suitable for TP Communication Compression?}
\label{sec:why fp8}
\subsection{Distribution of TP Intermediate Tensors}

We analyze the intermediate tensors involved in TP AllReduce (see Figure~\ref{fig:original_hist}). Our results indicate that these tensors are highly concentrated around zero while simultaneously exhibiting a long-tail distribution.
Let the TP intermediate tensor be $X = \{x_i\}_{i=1}^N$ with probability density $p_X(x)$, such that the zero-centered region satisfies $\int_{-\epsilon}^{\epsilon} p_X(x) \, dx \approx 1$, where $\epsilon \to 0^+$. The dense and long-tail subsets can be defined as $X_0 = \{ x_i \in X \mid |x_i| \le \epsilon \}$ and $X_L = \{ x_i \in X \mid |x_i| > \epsilon \}$, capturing the extremely small and relatively large values, respectively.
In other words, most values transmitted during TP communication are extremely small and densely clustered near zero, while only a few occupying the long tail. For such distributions, quantization must provide sufficient resolution around $X_0$; otherwise, small-magnitude values may collapse to the same quantization level, resulting in information loss.

\subsection{INT8 Incompatibility with TP Communication Compression}
\begin{figure}[t]
    \centering
    \includegraphics[width=1\linewidth]{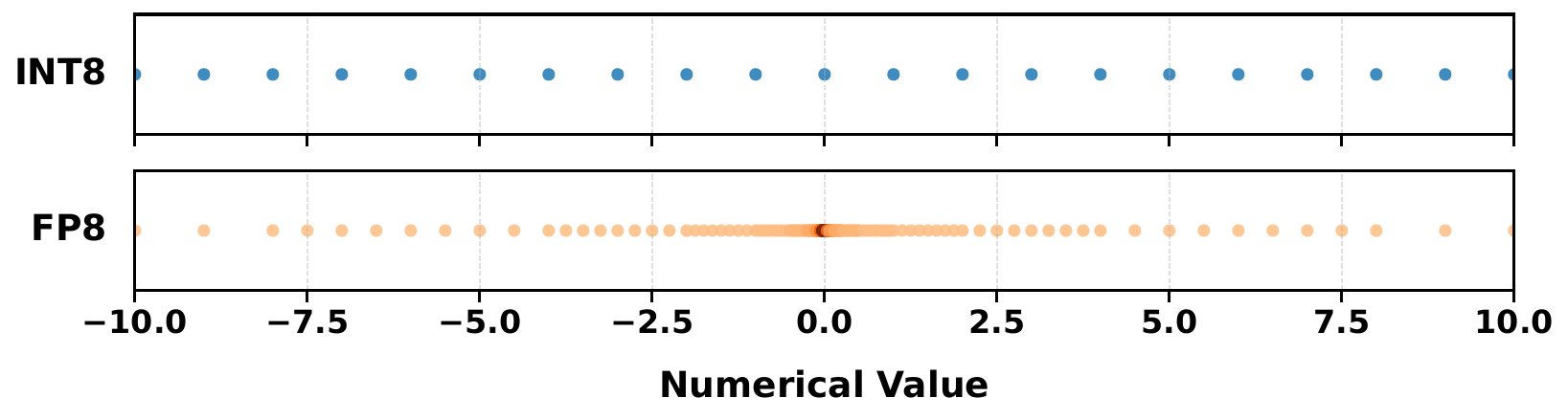}
    \caption{Data distribution characteristics of INT8 and FP8.}
    \Description{Distributions of values under INT8 and FP8 quantization. INT8 shows uniform quantization with evenly spaced bins, while FP8 exhibits non-uniform distribution with denser representation near zero and a wider dynamic range.}
    \label{fig:int8_fp8_plot}
\end{figure}

INT8 quantization typically adopts a fixed scale with zero-point, performing uniform quantization over the entire value range. However, this approach is unsuitable for TP intermediate tensors (see Figure~\ref{fig:original_hist}). INT8 maps the range using a fixed step size \(\Delta\), while TP intermediate tensors are highly concentrated near zero, forming a sharp peak. Figure~\ref{fig:int8_fp8_plot} shows the visual comparison of INT8 and FP8 representations. INT8 adopts uniform quantization with evenly spaced representable values, imposing the same quantization resolution on both high-density and low-density regions. As a result, the numerous small-magnitude values clustered around zero incur large relative quantization errors, leading to a pronounced degradation in training accuracy. Let the quantization error be
\begin{equation}
e_i = x_i - \hat{x}_i, \quad \hat{x}_i = Q_{\text{INT8}}(x_i), \quad i = 1, \dots, N.
\end{equation}
Due to the uniform step, multiple dense zero-centered values may map to the same integer, leading to collisions:
\begin{equation}
\exists x_i, x_j \in X_0, \quad Q_{\text{INT8}}(x_i) = Q_{\text{INT8}}(x_j), \quad i \neq j.
\end{equation}
Consequently, the mean squared error in the dense region remains significant, severely degrading training accuracy.
Our experiments on the tensors show that INT8 error is approximately uniformly distributed across the range (see Figure~\ref{fig:int8_fp8_error}), consistent with its uniform step mechanism. Therefore, INT8 can't provide fine-grained resolution in the dense zero region and is unsuitable for TP.

\subsection{The Mathematical Suitability of FP8}

FP8 addresses the limitations of INT8 by providing an exponentially scaled representation that achieves high precision near zero while maintaining a wide dynamic range. 
Figure~\ref{fig:int8_fp8_plot} illustrates the one-dimensional density of FP8 representable values, characterized by a pronounced concentration near zero and a long tail, which highlights FP8’s suitability for compressing TP intermediate tensors.
For values in the dense zero-centered region $X_0$, the quantization error is bounded by the unit in the last place (ULP):
\begin{equation}
|e_i^{\text{FP8}}| \le \text{ULP}(x_i) = 2^{\lfloor \log_2 |x_i| \rfloor - m}, \quad x_i \in X_0,
\end{equation}
where $m$ is the mantissa width. This implies that smaller values (lower exponent) are represented with finer precision and denser quantization points, closely matching the dense, small-magnitude peak observed in TP intermediate tensors.

\begin{figure}[t]
    \centering
    \includegraphics[width=0.9\linewidth]{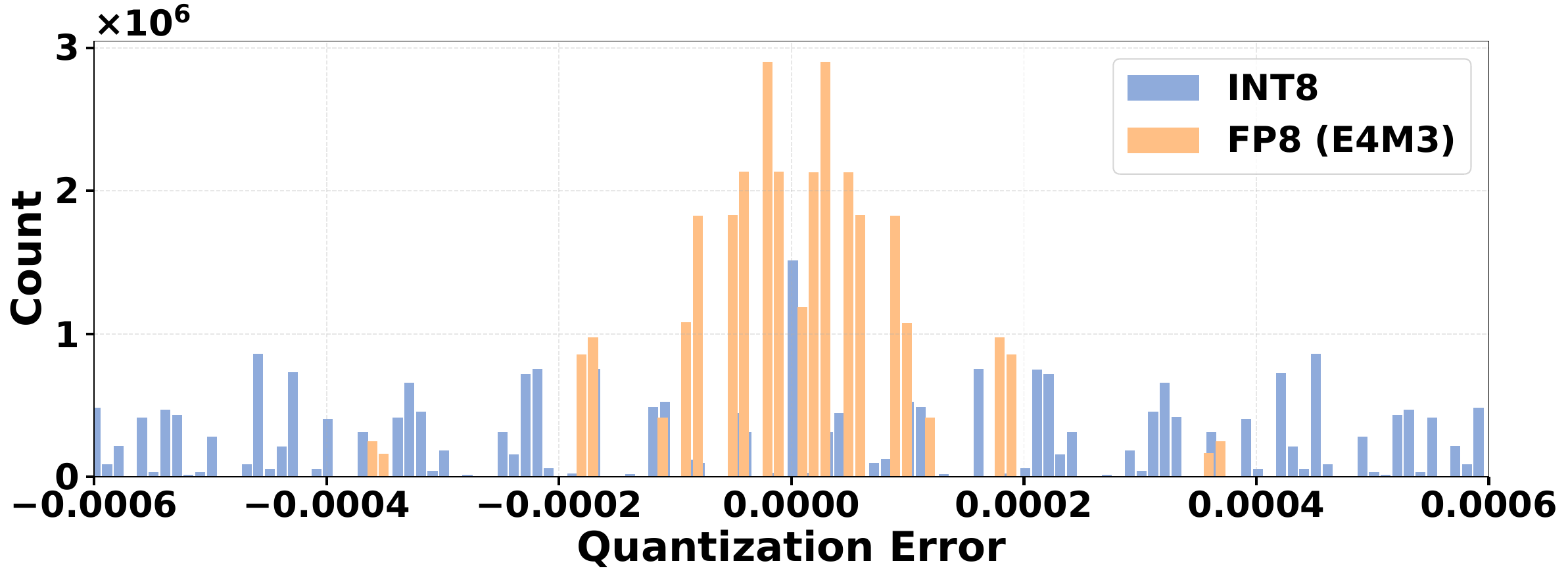}
    \caption{Quantization errors of INT8 and FP8.}
    \Description{Comparison of quantization errors between INT8 and FP8, showing how error magnitudes are distributed and highlighting differences in precision loss under the two formats.}
    \label{fig:int8_fp8_error}
\end{figure}

For large-magnitude values in the long-tail region $X_L$, FP8 step size grows exponentially, providing sufficient dynamic range. As a result, the overall tensor-level quantization error is significantly lower than INT8, while preserving high-fidelity representation near zero. This property can be approximated for E4M3 FP8 as
\begin{equation}
Q_{\text{FP8}}(x) \approx \text{sign}(x) \cdot 2^{E-B} \cdot \left(1 + M/2^m \right),
\end{equation}
where the exponent $E$ and mantissa $M$ are encoded with 4 and 3 bits, respectively. 
The exponential step ensures minimal error for critical small values, while still covering the long-tail region, achieving \textit{local high precision} and \textit{global dynamic range} simultaneously. 
The quantization errors of int8 and FP8 to TP intermediate tensor are as shown in Figure~\ref{fig:int8_fp8_error}, which demonstrates that the error of fp8 is obviously smaller than that of INT8. 
Overall, TP intermediate tensors exhibit distributions that are heavily skewed toward low-magnitude values, rendering uniform INT8 quantization susceptible to substantial precision loss. In contrast, FP8’s exponent-based representation provides finer granularity in the near-zero regime, making it more suitable for large-scale distributed training.

\section{Design of TACO}

\begin{figure*}[t]
    \centering
    \includegraphics[width=1\linewidth]{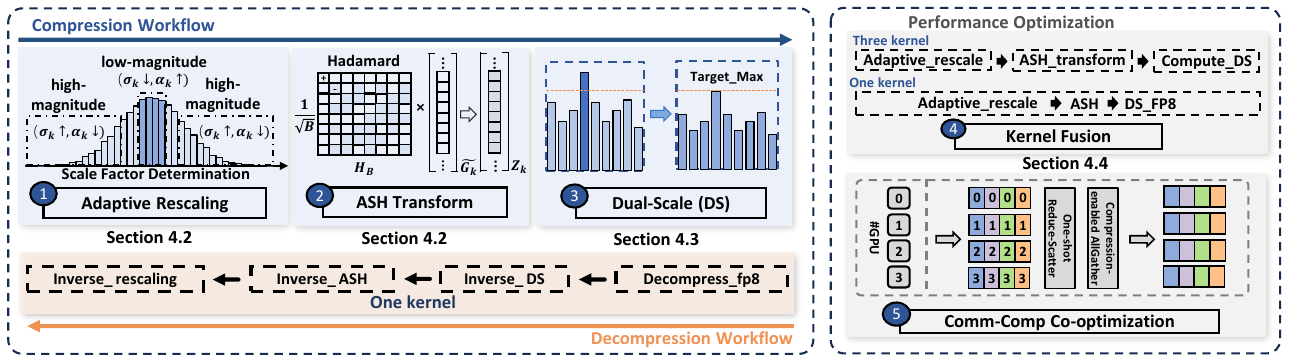}
    \caption{Overview of TACO. TP intermediate tensors are compressed via adaptive rescaling, the ASH transform, and dual-scale quantization, with kernel fusion and communication–computation co-optimization. Decompression is performed using a fused kernel for efficient reconstruction.}
    \Description{System overview of TACO. Tensor parallel intermediate tensors are first compressed using adaptive rescaling, the ASH transform, and dual-scale quantization. The system further applies kernel fusion and communication–computation co-optimization to improve efficiency. On the receiver side, a fused kernel is used for fast decompression and reconstruction of tensors.}
    \label{fig:taco_overview}
\end{figure*}

\subsection{High-Level Overview}
\label{subsec:overview}

Motivated by the observations above, we propose TACO. TACO specifically targets the compression of intermediate tensors exchanged in TP communication. It enhances the adaptability of FP8 quantization by dynamically regulating the energy distribution of these tensors, effectively mitigating the quantization error arising from their highly concentrated nature near zero. Furthermore, it enables high-throughput data transmission on GPU clusters through a unified kernel design. The overview of TACO is shown in Figure~\ref{fig:taco_overview}.

In the compression workflow, the limitations of standard Hadamard are first analyzed in Section~\ref{subsec:hadamard_limit}. To address the identified ``Zero-Collapse'' issue, the Adaptive Scale-Hadamard (ASH) transform is detailed in Section~\ref{subsec:ash}. This process involves calculating the second-order raw moment and scale factors (\ding{172}), followed by the ASH transform (\ding{173}). In Section~\ref{subsec:ds}, the Dual-Scale (DS) quantization (\ding{174}) is applied to map the transformed data to the FP8 format. Since executing these operations sequentially incurs significant memory overhead, kernel fusion (\ding{175}) is required to combine variance computation, ASH transform, and DS quantization into a single operator. The specific implementation and kernel fusion methods are detailed in Section~\ref{subsec:fusion}. Furthermore, to mitigate the overhead of compression latency, an overlap strategy (\ding{176}) is employed to parallelize computation and communication. The system performance optimization and overlap details are discussed in Section~\ref{subsec:overlap}.

In the decompression workflow, the data is first processed by the dequantization step, followed by the inverse DS quantization and inverse ASH transformation. We optimize the decompression process by fusing these operations into a unified \texttt{fused\_ash \_decompress\_kernel}. This fusion eliminates intermediate memory writes and reduces kernel launch overhead.

\subsection{Adaptive Scale--Hadamard Transform}
\label{subsec:ash}

\subsubsection{Motivation: Limitation of Standard Hadamard}
\label{subsec:hadamard_limit}

Despite the superior near-zero resolution of FP8 compared to INT8, its application to TP intermediate tensors is hindered by their extreme value concentration. We identify \textbf{``Zero-Collapse''} as the \textbf{root cause of failure}, where direct quantization or standard transformations cannot effectively map data into the representable range of FP8.

\textit{\textbf{Direct FP8 Quantization.}} Due to the lack of adaptive scaling, the vast majority of low-magnitude values fall into the subnormal range or underflow directly to zero, resulting in severe fidelity loss.

\textit{\textbf{Inefficacy of Standard Hadamard.}} While the standard Hadamard transform is effective for spreading outliers, it is fundamentally an \textit{isometric} transformation that preserves the Euclidean norm ($L_2$ energy) of the input vector. Consequently, data blocks with inherently low energy remain confined to a narrow numerical range even after rotation, failing to occupy the effective bits of FP8.

As illustrated in Figure~\ref{fig:hadamard_distribution}, the distribution of the transformed data remains sharply peaked around zero, indicating that the standard Hadamard transform fails to sufficiently disperse these dense, low-magnitude clusters. Consequently, this results in persistent underutilization of FP8’s high-precision dynamic range.

\subsubsection{ASH Transform}
\label{subsubsec:ash}

\begin{figure}[t]
    \centering
    \includegraphics[width=1.0\linewidth]{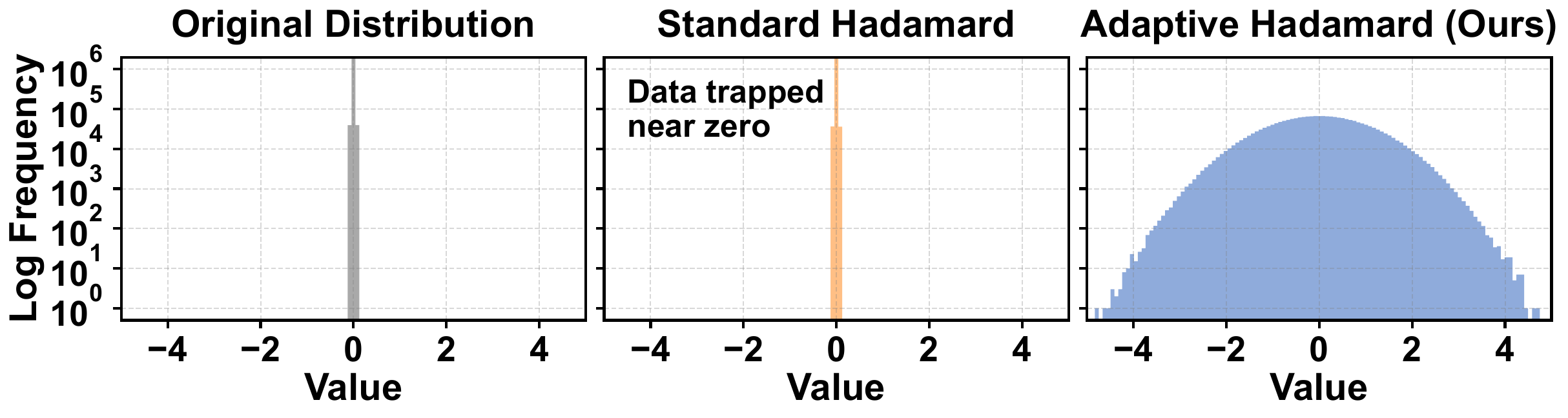}
    \caption{Distribution of tensors before and after Hadamard-based transformations.
AS-Hadamard redistributes densely clustered near-zero values into FP8’s high-precision quantization range, unlike the standard Hadamard transform.}
    \Description{Comparison of tensor distributions before and after Hadamard-based transformations. The AS-Hadamard transform redistributes values that are densely concentrated near zero into a more uniform distribution, better matching FP8 quantization precision, whereas the standard Hadamard transform does not achieve this effect as effectively.}
    \label{fig:hadamard_distribution}
\end{figure}

To overcome the limitations of the standard Hadamard transform and prevent zero-collapse, we propose the Adaptive Scale--Hadamard Transform, which combines block-wise energy rescaling with orthogonal Hadamard rotation for high-fidelity quantization. This design amplifies low-magnitude blocks while appropriately scaling high-magnitude blocks, enabling the transformed data to fully exploit FP8’s representable range.
Let $\mathbf{X} \in \mathbb{R}^{N}$ denote the flattened TP intermediate tensor prior to transfer. We partition $\mathbf{X}$ into $M$ contiguous blocks of size $B$:
\begin{equation}
    \mathbf{X} = [G_1, G_2, \dots, G_M], \quad G_k \in \mathbb{R}^B,
    \label{eq:ash_partition}
\end{equation}
where $B$ is chosen to fit entirely in GPU shared memory for low-latency, block-local processing.
The effect of ASH transform is clearly visualized in Figure~\ref{fig:hadamard_distribution}. Compared to the standard Hadamard transform, ASH transform effectively disperses the dense near-zero clusters, expanding them into FP8's high-precision quantization range and yielding a more balanced distribution.

\paragraph{\textbf{Block-wise Adaptive Rescaling.}}
To explicitly control the numerical magnitude of each block, ASH begins with block-wise adaptive energy rescaling. We specifically adopt the second-order raw moment (energy) rather than variance to estimate the local scale. This is motivated by the observation that TP intermediate tensors are typically zero-centered; thus, the raw energy captures the effective signal magnitude required for quantization without incurring the computational overhead of mean centering. For block $G_k$, we calculate its root mean square (RMS) amplitude as follows:
\begin{equation}
    \sigma_k = \sqrt{\frac{1}{B} \sum_{j=1}^{B} G_{k,j}^2 + \epsilon} ,
    \label{eq:ash_sigma}
\end{equation}
where $\epsilon$ is a small constant ensuring numerical stability for near-zero blocks.  
We then compute an adaptive scaling factor to map each block's energy to a reference target level:
\begin{equation}
    \alpha_k = \frac{\tau}{\sigma_k} ,
    \label{eq:ash_alpha}
\end{equation}
where $\tau$ denotes a constant target energy aligned with FP8’s effective dynamic range. This scaling strategy amplifies low-magnitude blocks while attenuating higher-magnitude ones, effectively normalizing the dynamic range across all blocks. The rescaled block is computed element-wise as $\tilde{G}_k = \alpha_k \cdot G_k$, where each block $G_k$ has its own adaptive scale $\alpha_k$. This ensures that every block is normalized independently, providing distinguishable numerical ranges across blocks before the rotation stage.
By doing so, we effectively maximize the utilization of FP8's high-precision range and minimize quantization errors for near-zero values. Importantly, this computation involves only lightweight reductions and element-wise multiplications, making it highly efficient and naturally suited for massively parallel GPU execution, where thousands of blocks can be processed concurrently with minimal overhead.

\paragraph{\textbf{Orthogonal Hadamard Rotation.}}
After rescaling, each block undergoes an \textbf{orthogonal} Walsh--Hadamard transform:
\begin{equation}
    \mathbf{Z}_k = \frac{1}{\sqrt{B}} \mathbf{H}_B \tilde{G}_k ,
    \label{eq:ash_rotation}
\end{equation}
where $\mathbf{H}_B$ denotes the Hadamard matrix of order $B$. The factor $1/\sqrt{B}$ is incorporated to explicitly enforce orthogonality, thereby preserving the energy of the transformation. Since the normalized Hadamard matrix is bothsymmetric and orthogonal ($\mathbf{H}_B^T = \mathbf{H}_B^{-1} = \frac{1}{\sqrt{B}}\mathbf{H}_B$), the transformation is exactly invertible.

Crucially, because the rotation is energy-preserving, it redistributes the dense low-magnitude clusters without disrupting the block-wise scaling established during the rescaling step. The transformed data exhibits an approximate zero-mean, Gaussian-like distribution, which naturally aligns with the non-uniform exponent density of FP8. As illustrated in Figure~\ref{fig:hadamard_distribution}, ASH significantly improves the utilization of FP8's effective dynamic range compared to the standard Hadamard transform. For efficiency, this operation is implemented using the Fast Walsh--Hadamard Transform (FWHT) in an in-place manner within shared memory, reducing computational complexity from \textbf{$O(B^2)$} to \textbf{$O(B \log B)$}.

\subsection{\textbf{Dual-Scale FP8 Quantization}}
\label{subsec:ds}

Following the ASH transform, the rotated tensor $\mathbf{Z}_k$ exhibits a more favorable Gaussian-like distribution. Nevertheless, careful scale management remains critical due to the rigid upper bound of the FP8 format. Without proper scaling, high-magnitude values can exceed the maximum representable range ($Q_{\max}$), causing overflow or severe saturation. Such numerical instability is a primary contributor to training divergence and abrupt loss spikes.

\paragraph{Post-Rotation Quantization Scale.}
To ensure that all values strictly fit in the FP8 representable range, we compute a block-wise post-rotation scale based on the maximum absolute value in the block:
\begin{equation}
    s_k = \frac{\max(|\mathbf{Z}_k|)}{Q_{\max}} ,
    \label{eq:post_scale}
\end{equation}
where $\max(|\mathbf{Z}_k|)$ denotes the maximum value in block $k$, and $Q_{\max}$ is the largest representable FP8 value. 
The data are then quantized element-wise to FP8:
\begin{equation}
    \mathbf{q}_k = \mathrm{Cvt}_{\mathrm{FP8}}\!\left(\frac{\mathbf{Z}_k}{s_k}\right) ,
    \label{eq:quantization}
\end{equation}
where $\mathrm{Cvt}_{\mathrm{FP8}}$ denotes the intrinsic conversion function (NVIDIA's \texttt{\_nv\_cvt\_float\_to\_fp8}). By strictly enforcing this mapping, numerical overflow is effectively prevented, and the available bit-width is maximally and efficiently utilized throughout computation.

\begin{algorithm}[t]
\caption{TACO: Tensor-Parallel Adaptive Communication Compression}
\label{alg:taco}
\begin{algorithmic}[1]

\Require Input intermediate tensor $\mathbf{X}$; block size $B$; target energy $\tau$; FP8 maximum value $Q_{\max}$
\Ensure Reconstructed intermediate tensor $\mathbf{X}'$

\Statex \textbf{Sender-side: Fused Compression Kernel}

\State Partition $\mathbf{X}$ into $M$ contiguous blocks $\{G_1, G_2, \dots, G_M\}$ of size $B$
\For{each block $G_k$ \textbf{in parallel}}
    \State Load $G_k$ into shared memory / registers

    \State \textbf{Block-wise Adaptive Rescaling}
    \State $\sigma_k \gets \sqrt{\frac{1}{B} \sum_{j=1}^{B} G_{k,j}^2 + \epsilon}$
    \State $\alpha_k \gets \tau / \sigma_k$
    \State $\tilde{G}_k \gets \alpha_k \cdot G_k$

    \State \textbf{Orthogonal Hadamard Rotation}
    \State $\mathbf{Z}_k \gets \frac{1}{\sqrt{B}} \mathrm{FWHT}(\tilde{G}_k)$

    \State \textbf{Post-Rotation FP8 Quantization}
    \State $s_k \gets \max(|\mathbf{Z}_k|) / Q_{\max}$
    \State $\mathbf{q}_k \gets \mathrm{Cvt}_{\mathrm{FP8}}(\mathbf{Z}_k / s_k)$
\EndFor

\State Communicate $\{(\mathbf{q}_k, \alpha_k, s_k)\}_{k=1}^{M}$ using TP collectives

\Statex \textbf{Receiver-side: Decompression Kernel}

\For{each received tuple $(\mathbf{q}_k, \alpha_k, s_k)$ \textbf{in parallel}}
    \State $\hat{\mathbf{Z}}_k \gets \mathrm{Cvt}_{\mathrm{FP32}}(\mathbf{q}_k) \cdot s_k$
    \State $\hat{G}_k \gets \frac{1}{\sqrt{B}} \mathrm{FWHT}(\hat{\mathbf{Z}}_k)$
    \State $G'_k \gets \hat{G}_k / \alpha_k$
    \State Append $G'_k$ to $\mathbf{X}'$
\EndFor

\State \Return $\mathbf{X}'$

\end{algorithmic}
\end{algorithm}

\paragraph{\textbf{Dual-Scale Reconstruction.}}
To enable high-fidelity reconstruction, TACO utilizes two distinct scalars per block to separately control the data distribution and the quantization range. Both the adaptive rescaling factor $\alpha_k$ (Eq.~\ref{eq:ash_alpha}), which normalizes block energy, and the quantization scale $s_k$ (Eq.~\ref{eq:post_scale}), which prevents numerical overflow, are transmitted alongside the compressed tensor. Reconstruction strictly follows the reverse order of the compression sequence:
\begin{align}
    \hat{\mathbf{Z}}_k &= \mathrm{Cvt}_{\mathrm{FP32}}(\mathbf{q}_k) \cdot s_k , \label{eq:recon_dequant} \\
    \hat{G}_k &= \frac{1}{\sqrt{B}}\mathbf{H}_B \hat{\mathbf{Z}}_k , \label{eq:recon_ifwht} \\
    G'_k &= \hat{G}_k / \alpha_k . \label{eq:recon_rescale}
\end{align}
Specifically, Eq.~\ref{eq:recon_dequant} recovers the post-rotation magnitude, Eq.~\ref{eq:recon_ifwht} applies the inverse orthogonal Hadamard transform to restore the original ordering of the data, and Eq.~\ref{eq:recon_rescale} reverses the adaptive rescaling.
By decoupling these two factors, this mechanism mitigates the conflict between resolving small values and containing large values. It ensures that low-magnitude clusters are safeguarded against underflow, while high-magnitude blocks remain strictly bounded within the FP8 representable range throughout training.

The communication overhead of transmitting two scalars per block is negligible compared to the substantial bandwidth savings achieved by FP8 compression. Despite this minimal overhead, the additional scaling metadata is critically important for maintaining numerical stability and ensuring stable training convergence. As a result, TACO enables highly efficient FP8-based TP communication, while having minimal impact on overall training accuracy.

\subsection{Performance Optimization}

\subsubsection{System-Level Kernel Fusion Optimizations.}
\label{subsec:fusion}
\begin{figure}[t]
    \centering
    \includegraphics[width=1.0\linewidth]{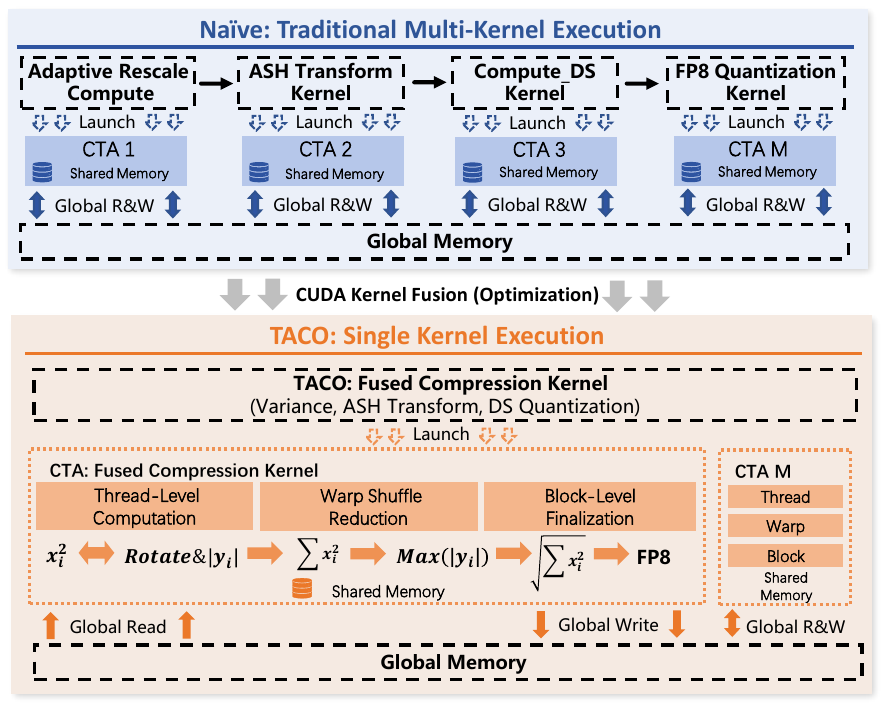}
    \caption{Comparison of Na\"ive Multi-Kernel Execution vs. TACO Single Kernel Execution. The upper part illustrates the high memory traffic in traditional methods, while the lower part demonstrates the efficiency of fused kernel execution with parallel CTAs.}
    \Description{Comparison between naive multi-kernel execution and TACO single-kernel execution. The naive approach requires multiple kernel launches and incurs high memory traffic due to repeated intermediate data movement. In contrast, TACO fuses operations into a single kernel with parallel CTAs, significantly reducing memory traffic and improving execution efficiency.}
    \label{fig:taco_fusion}
\end{figure}

TACO implements a fully fused compression kernel (Algorithm~\ref{alg:taco}) that integrates adaptive rescale factor computation, the ASH transform, and DS quantization into a single GPU kernel (Figure~\ref{fig:taco_fusion}). By eliminating redundant global memory accesses and kernel launches, this design significantly increases arithmetic intensity and improves overall efficiency. In contrast, naïve implementations typically launch separate reduction kernels to compute block-wise variance for adaptive energy normalization and the post-rotation maximum magnitude for FP8 scaling, incurring substantial overhead.

TACO effectively eliminates this redundancy by coalescing both reductions within a single fused kernel using warp-level primitives and shared memory.
Concretely, each thread locally computes both the squared input value for variance estimation and the absolute value of its rotated output for post-rotation scaling.
Warp-level shuffle operations are then used to simultaneously aggregate partial sum and partial maximum with minimal synchronization overhead, after which shared memory efficiently finalizes the block-level reductions to produce $\sigma_k$ and $s_k$ without launching auxiliary kernels. This design significantly reduces global memory traffic and kernel launch overhead, enabling fully block-local computation for both normalization and quantization scale derivation.

\subsubsection{System performance optimization}

\label{subsec:overlap}
To better integrate TACO with communication protocols, we implement two optimizations: refining buffer distribution and integrating TACO with COCCL, a compressible collective communication library, followed by performance tuning, ensuring efficient TP communication across GPUs, which allows TACO to be seamlessly embedded into existing collective primitives with minimal protocol modification.

First, each compressed block requires two scalar parameters: the adaptive pre-scaling factor $\alpha_k$ and the post-rotation quantization scale $s_k$. TACO stores these scalars contiguously with the corresponding FP8-compressed payload in global memory, enabling a zero-copy metadata layout.
During decompression, the kernel retrieves $\alpha_k$ and $s_k$ via simple pointer arithmetic, avoiding explicit metadata copies as well as additional communication launches for scale gathering.
Moreover, this layout enables fully coalesced global memory accesses during reconstruction, as threads within a block read contiguous FP8 values followed immediately by their associated metadata,further minimizing memory latency.

\begin{figure}[t]
    \centering
    % width=\linewidth 确保图片自动缩放以适应行宽
    \includegraphics[width=\linewidth]{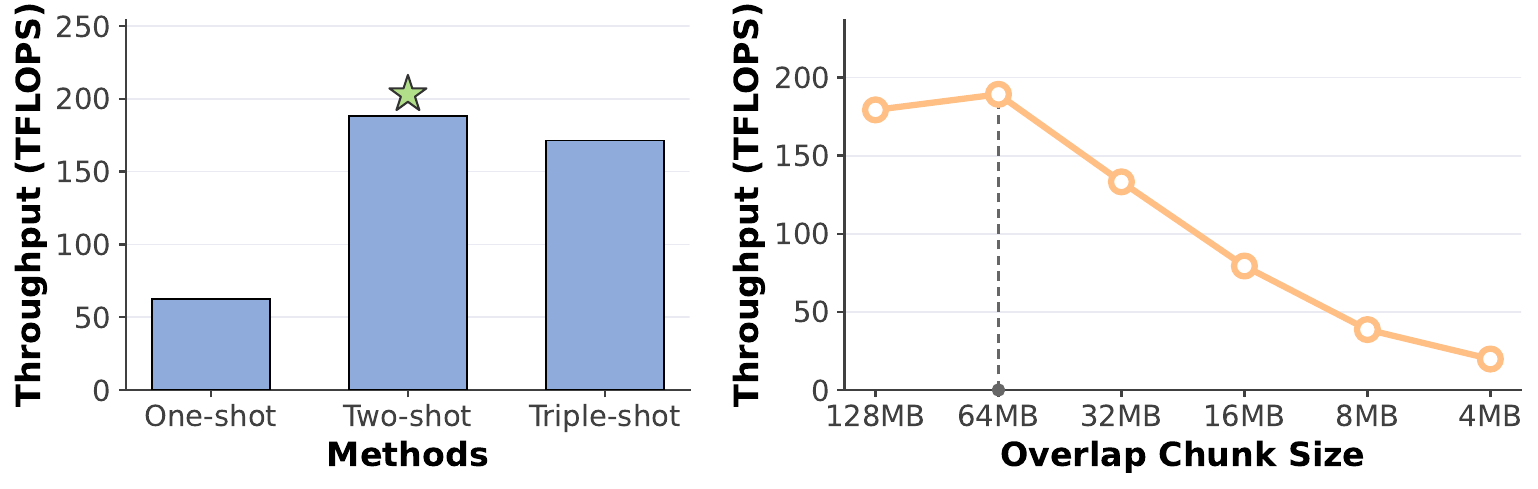}
    \caption{Throughput comparison. Left: Performance of different shot methods. Right: Impact of overlap chunk size on throughput.}
    \Description{Throughput comparison across different configurations. The left plot evaluates the performance of different shot methods, while the right plot analyzes the impact of overlap chunk size on throughput, demonstrating the trade-off between communication-computation overlap and execution efficiency.}
    \label{fig:throughput_chart}
\end{figure}

Second, integrating TACO into traditional ring- or tree-based communication algorithms results in frequent compression within a single communication, leading to significant performance degradation and accumulated compression errors. To reduce compression overhead, we integrate TACO into COCCL~\cite{coccl}, a compressible collective communication protocol built on NCCL.
We tune this library as shown in Figure~\ref{fig:throughput_chart} and find that, TP communication scenario—where all communication occurs within a node—using COCCL’s two-shot AllReduce as the communication algorithm achieves the best overall performance. The two-shot algorithm decomposes AllReduce into ReduceScatter and AllGather. The ReduceScatter phase consists of one compressed AlltoAll operation followed by a single local reduction.
This design effectively reduces TACO execution to two operations per communication round, minimizing compression frequency and increasing compression block granularity. As a result, it significantly lowers execution overhead when integrating TACO with communication while preserving accuracy. In addition, COCCL incorporates a two-level overlap strategy. Through optimization of the overlap granularity, we determine that overlapping data blocks of size 64 MB yields optimal performance, further masking the computational latency of TACO.

\section{Evaluation}

\subsection{Experimental Setup} 

\textit{\textbf{Platforms.}} We evaluate TACO on a GPU-centric cluster comprising two nodes, each equipped with two Intel XEON(R) Platinum 8558 CPUs (192 cores) and eight NVIDIA H100 SXM5 GPUs with 80 GB memory. The nodes are interconnected via four 400Gbps InfiniBand links, providing an aggregate inter-node bandwidth of 1.6 Tbps. The software stack includes CUDA~12.6 and NVIDIA driver~550.90.07.

\textit{\textbf{Baselines.}}
We compare TACO against the following communication strategies:
Baseline (w/o Comp), which performs standard distributed training without communication compression;
TahQuant~\cite{he2025tahquant}, a PP compression method serving as a representative baseline for communication quantization;
SDP4bit~\cite{jia2024sdp4bit}, a 4-bit quantization method optimized for DP gradient compression.

\textit{\textbf{Models and Datasets.}}
In Section~\ref{sec:acc}, we evaluate compression algorithms on GPT-350M trained on the Pile dataset~\cite{gao2020thepile}, using a learning rate schedule of $(3\times10^{-4} \rightarrow 3\times10^{-5})$, with a global batch size of 256 and 10,000 training iterations. We further extend the evaluation to a larger model, GPT-6.7B, trained on the same Pile dataset. In addition, Qwen2.5-7B~\cite{qwen2.5} is trained on the Open-Web-Math dataset~\cite{paster2023openwebmath}, following a learning rate schedule of $(3\times10^{-4} \rightarrow 3\times10^{-5})$, with a global batch size of 64 for 10,000 iterations, to evaluate the generalization ability of the proposed method across different model families and data distributions.
In Section~\ref{sec:3d_system_eval}, we conduct large-scale evaluations under a 3D parallel training configuration, where GPT-6.7B is trained from scratch on the Pile dataset using PyTorch~\cite{paszke2019pytorch} v2.5.1 and Megatron-LM~\cite{shoeybi2019megatronlm}, with parallelism configured as (TP~=~4, PP~=~2, DP~=~2). 

\textit{\textbf{Metrics.}} We report End-to-end throughput (TFLOPS) to measure efficiency, and Model quality (Validation/Test Loss) to evaluate convergence. Degradation (Deg.) is reported as the relative percentage increase in loss relative to the BF16 baseline.

\begin{table}[t]
\centering
\caption{Accuracy comparison under TP=8 after 10{,}000 training iterations for the BF16 baseline, TahQuant, and TACO; degradation (Deg.) indicates the relative loss increase over the baseline.}
\label{tab:tp8_accuracy}
\resizebox{\columnwidth}{!}{%
\begin{tabular}{lcccc}
\toprule
Method 
& Val Loss $\downarrow$ 
& Test Loss $\downarrow$ 
& Val Deg.  $\downarrow$ 
& Test Deg.  $\downarrow$ \\
\midrule
Baseline & 2.389899 & 2.344701 & -- & -- \\
TahQuant & 2.458742 & 2.413642 & +2.88 \% & +2.94 \%\\
\textbf{TACO} & \textbf{2.395784} & \textbf{2.351210} & \textbf{+0.25 \%} & \textbf{+0.28 \%}\\
\bottomrule
\end{tabular}
}
\end{table}

\subsection{Evaluation of Accuracy with TP}
\label{sec:acc}

In this section, we systematically evaluate the impact of communication compression on model convergence under TP settings, where intermediate tensors are exchanged across GPUs during both forward and backward passes. First, we benchmark the overall \textit{End-to-End Performance} in Section~\ref{sec:end_to_end} under high-parallelism configurations (up to TP8) to comprehensively demonstrate robustness and convergence stability at scale. 
Next, we perform a \textit{Component-wise Analysis} in Section~\ref{sec:component} of TACO’s internal mechanisms—ASH and DS—to assess their contributions to numerical stability and precision recovery. 
We further extend this evaluation in Section~\ref{subsubsec:large_scale_accuracy} to large-scale models (GPT-6.7B and Qwen-2.5-7B), rigorously verifying that TACO maintains stable optimization dynamics and exhibits negligible accuracy degradation under aggressive TP compression.

\subsubsection{End-to-End Convergence Comparison}
\label{sec:end_to_end}

To evaluate the effectiveness of the proposed TACO framework for compressing TP intermediate tensors, we benchmark its end-to-end training accuracy against the state-of-the-art communication compression method TahQuant under a TP8 configuration. Table~\ref{tab:tp8_accuracy} reports the final validation and test losses for the uncompressed BF16 baseline, TahQuant, and TACO.
After 10{,}000 training iterations, the BF16 baseline achieves a validation loss of 2.389899 and a test loss of 2.344701, serving as the reference for assessing compression-induced degradation. While TahQuant reduces communication volume, it incurs a substantial accuracy penalty, with validation and test losses increasing by $+2.88\%$ and $+2.94\%$, respectively. This degradation suggests that quantization errors introduced into TP intermediate tensors accumulate across layers and iterations, ultimately impeding convergence toward the full-precision optimum.

In contrast, TACO exhibits consistently near-lossless accuracy. As shown in Table~\ref{tab:tp8_accuracy}, TACO attains a validation loss of 2.395784 and a test loss of 2.351210, corresponding to only $+0.25\%$ and $+0.28\%$ degradation relative to BF16. Compared to TahQuant, TACO improves fidelity by more than an order of magnitude. This robustness is particularly significant in TP settings, where intermediate tensors are exchanged at every layer and iteration, and even small numerical perturbations can rapidly propagate and destabilize training.

\begin{figure}[t]
    \centering
    \includegraphics[width=1\linewidth]{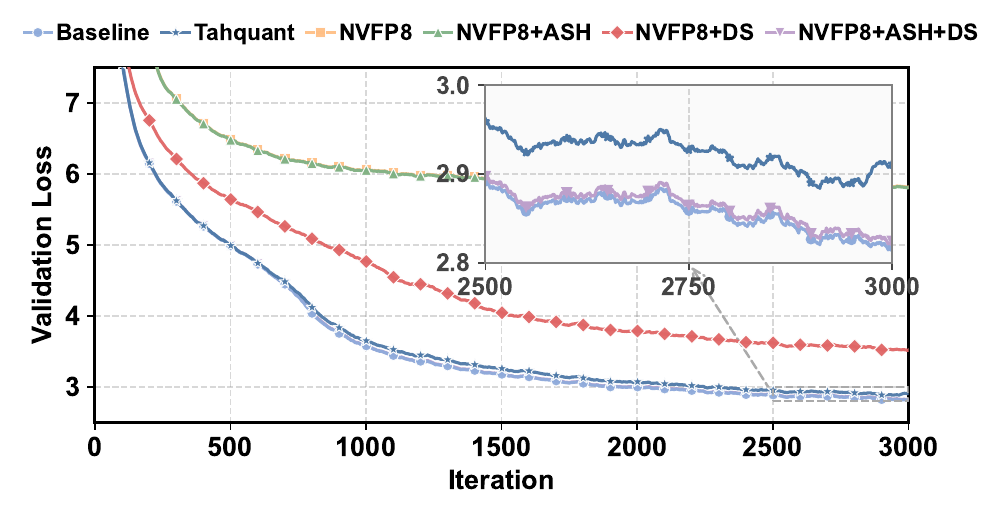}
    \caption{Ablation of TACO components on convergence stability. The plot demonstrates the progressive stability gained by combining ASH and DS, with the final framework (purple) nearly overlapping the uncompressed baseline (blue).}
    \Description{Ablation results on convergence stability. The study evaluates the effect of ASH and DS components individually and in combination. As components are added, training stability improves progressively, and the full TACO system closely matches the uncompressed baseline, demonstrating negligible impact on convergence.}
    \label{fig:loss(ash_ds)}
\end{figure}

\subsubsection{Component-wise Analysis}
\label{sec:component}

To elucidate how TACO preserves training stability, we systematically analyze its core architectural components: \textit{ASH} and \textit{DS}. Figure~\ref{fig:loss(ash_ds)} presents the convergence of these configurations under TP4.
The empirical evidence reveals that naively applying standard NVFP8 compression to TP intermediate tensors leads to immediate and catastrophic divergence, with the validation loss spiking to 5.605692. As shown by the NVFP8 curve in Figure~\ref{fig:loss(ash_ds)}, the loss quickly plateaus at an extremely high value (near 6.0) and completely fails to track the downward trend of the baseline. This indicates that standard 8-bit quantization without conditioning is clearly insufficient for the precision requirements of TP. This failure is rooted in the highly non-uniform nature of the numerical distributions within these tensors. TP intermediate tensors typically exhibit a dense clustering of values near zero; such a distribution fails to utilize the discrete representable points of the FP8 format effectively, resulting in significant quantization noise that destabilizes the gradient flow across parallel partitions.

Integrating Dual-Scale (DS) quantization in isolation provides only partial stabilization of the training objective, reducing the validation loss to 3.300491. By partitioning TP intermediate tensors into finer sub-blocks and assigning independent scaling factors, DS mitigates precision loss caused by local dynamic range mismatches. However, the training trajectory (red diamonds in Figure~\ref{fig:loss(ash_ds)}) remains substantially above the uncompressed baseline, indicating that DS alone cannot fully recover full-precision performance. Without prior reshaping of the numerical distribution, the FP8 mantissa bits remain underutilized for the majority of densely clustered values, resulting in persistent information loss that limits convergence.

Crucially, the full TACO configuration (NVFP8 + ASH + DS) achieves near-baseline performance, with a validation loss of 2.667557. As shown by the purple trajectory in Figure~\ref{fig:loss(ash_ds)}, this configuration closely tracks the baseline curve, particularly during the later stages of training, as highlighted in the magnified inset, and consistently outperforms TahQuant under the same setting. In this synergy, ASH acts as a preconditioner that disperses dense clusters and flattens the distribution, while DS explicitly aligns the transformed blocks to the FP8 representable range. Notably, ASH alone yields limited improvement, since transformed values can still exceed the FP8 range without DS’s adaptive scaling. These results demonstrate that the combination of ASH and DS is essential for enabling high-fidelity, low-bit communication in TP training.

\begin{figure}[t]
    \centering
    % Figure: Component Ablation
    \includegraphics[width=1.0\linewidth]{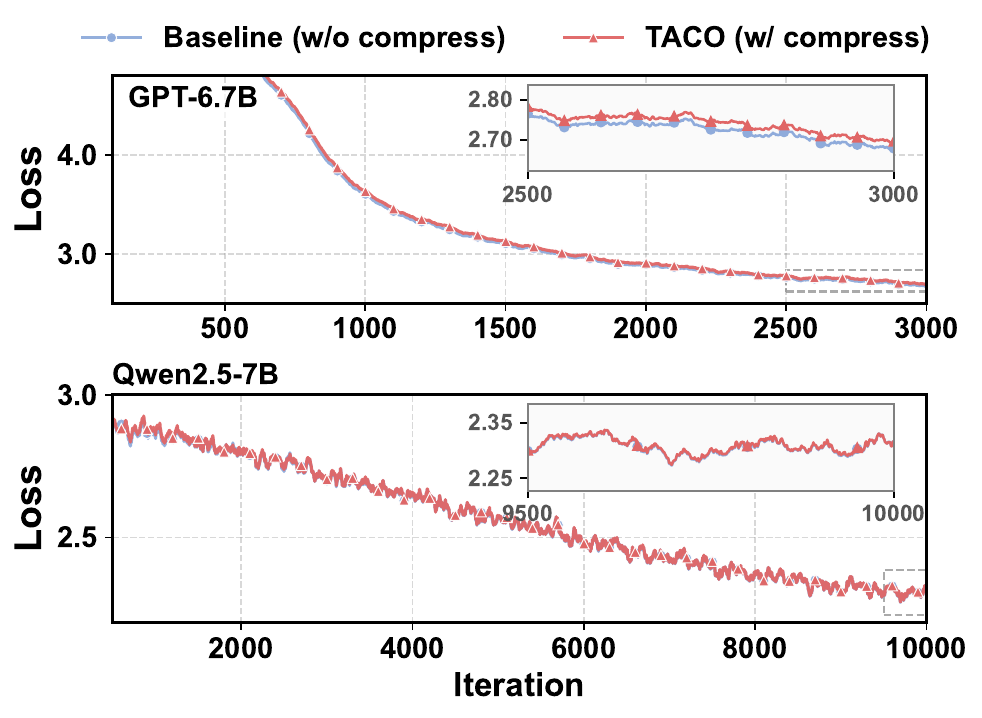}
    \caption{Validation loss comparison between the baseline (no compression) and TACO on GPT 6.7B and Qwen2.5-7B.}
    \Description{Validation loss comparison between the BF16 baseline (no compression) and TACO on GPT 6.7B and Qwen2.5 7B.}
    \label{fig:gpt6.7B_qwen7B}
\end{figure}

\subsubsection{Large-Scale Model Verification}
\label{subsubsec:large_scale_accuracy}

To evaluate the scalability and numerical robustness of the proposed framework, we further conduct experiments on two representative large-scale language models, GPT 6.7B and Qwen-2.5 7B. As shown in Figure~\ref{fig:gpt6.7B_qwen7B}, we compare the training loss trajectories between baseline (no compression) and TACO. For GPT 6.7B, TACO achieves a final validation loss of 2.570587, compared to 2.552718 for baseline, corresponding to a marginal degradation of +0.70\%. Similarly, on Qwen-2.5 7B, TACO reaches a loss of 2.257332 versus 2.256751 for the baseline, resulting in an extremely small degradation of +0.03\%.

Across both model families, the loss curves remain stable throughout training, and the performance gap between TACO and full-precision training is negligible. These results demonstrate that TACO consistently preserves optimization stability under aggressive communication compression, and generalizes well across different architectures and scales without requiring additional tuning.

\subsection{Ablation Study}
\label{sec:Ablation_Study}

We systematically investigate the impact of Hadamard-based transform and the selection of low-bit formats on training convergence under TP4. In Section~\ref{secsec:Component_Ablation}, we analyze the effectiveness of ASH by comparing it against standard Hadamard transform. In Section~\ref{secsec:Format_Ablation}, we evaluate the sensitivity to different quantization formats. 

\begin{figure}[t]
    \centering
    % Figure: Component Ablation
    \includegraphics[width=1.0\linewidth]{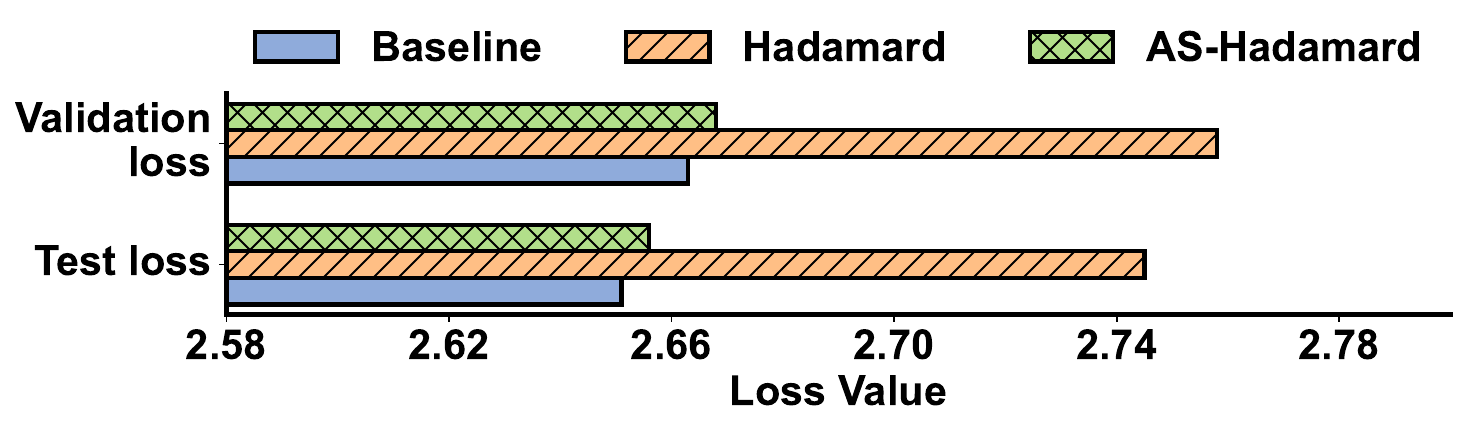}
    \caption{Validation and test loss for baseline, standard Hadamard, and ASH under TP4.}
    \Description{Ablation study under TP4 comparing baseline, standard Hadamard, and ASH. Standard Hadamard degrades both validation and test performance, while ASH mitigates this issue and restores accuracy close to the uncompressed baseline, demonstrating its effectiveness in preserving training quality.}
    \label{fig:tp_ablation_component}
\end{figure}

\subsubsection{Component Ablation.}
\label{secsec:Component_Ablation}

Our analysis compares the uncompressed baseline against the standard Hadamard transform and the proposed ASH mechanism. Applying a standard Hadamard transform increases the validation loss from 2.663061 to 2.757688, corresponding to a $+3.55\%$ degradation. This accuracy drop occurs because the standard rotation does not sufficiently recondition the dense clusters of near-zero values inherent in TP intermediate tensors. As a result, a large fraction of values remains concentrated in a narrow range, severely underutilizing the FP8 mantissa. In contrast, ASH restores performance to near-baseline levels, achieving a validation loss of 2.667557 (a marginal $+0.17\%$ degradation) by spreading values more uniformly across the representable range, thereby minimizing quantization error for small-magnitude elements.
As illustrated in Figure~\ref{fig:tp_ablation_component}, the standard Hadamard transform exhibits a visible accuracy gap, whereas ASH effectively recovers the lost fidelity. These results confirm that adaptive distribution reshaping is a prerequisite for high-precision, low-bit communication.

\subsubsection{Format Ablation.}
\label{secsec:Format_Ablation}

\begin{figure}[t]
    \centering
    % Figure: Format Ablation
    \includegraphics[width=1\linewidth]{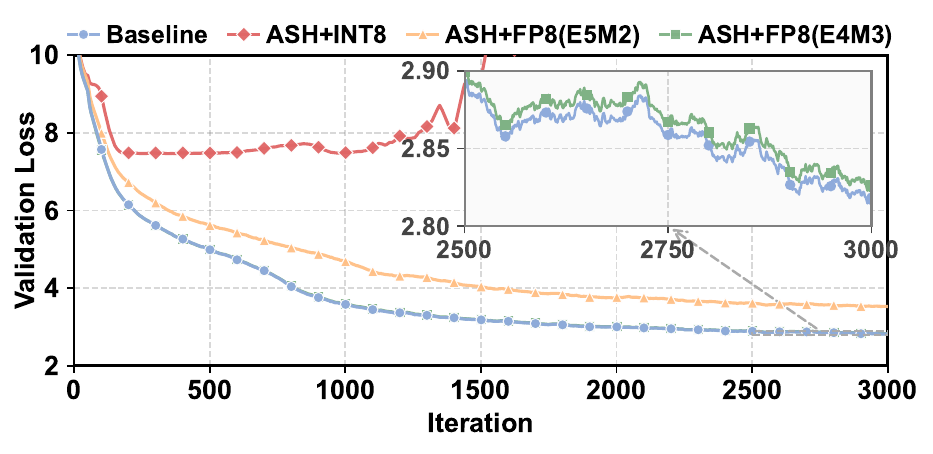}
    \caption{Validation loss curves for ASH combined with different low-bit formats. INT8 causes complete divergence}
    \Description{Ablation study of ASH with different low-bit formats. The results show that ASH is stable under FP8-based settings, while INT8 leads to complete divergence during training, demonstrating that aggressive quantization without sufficient representational range is incompatible with the proposed framework.}
    \label{fig:tp_ablation_format}
\end{figure}

We further evaluate the effectiveness of different low-bit numerical formats when combined with ASH, with results shown in Figure~\ref{fig:tp_ablation_format}. The choice of numerical format proves critical to training stability, as reflected by the markedly different loss trajectories observed in our ablation study. ASH+INT8 denotes INT8 quantization applied after ASH transform. As illustrated in Figure~\ref{fig:tp_ablation_format}, this configuration leads to catastrophic divergence. Although the loss remains superficially stable during the initial $\sim$1,500 iterations, it subsequently exhibits a sharp exponential increase, ultimately reaching a validation loss of 68.10. This failure arises because INT8’s limited  range cannot accommodate the broadened  tensor distribution produced by ASH, resulting in saturation of high-magnitude values and collapse of small-magnitude ones.

The floating-point formats offer significantly better robustness. The FP8 (E5M2) configuration partially mitigates the divergence seen in INT8, maintaining a stable downward trend. However, as shown in the magnified inset of Figure~\ref{fig:tp_ablation_format}, the E5M2 curve (orange triangles) remains consistently higher than the baseline, settling at a validation loss of 3.305199. This $+24.1\%$ degradation indicates that while the E5M2 format provides sufficient dynamic range (5 bits for exponent), its limited 2-bit mantissa lacks the necessary resolution to represent the critical details of the reshaped tensors.

In contrast, the FP8 (E4M3) format achieves the optimal balance between range and precision. As illustrated in Figure~\ref{fig:tp_ablation_format}, the E4M3 curve (green squares) tracks the uncompressed baseline (blue circles) with remarkable fidelity throughout the entire training process, achieving a near-lossless validation loss of 2.667557, only a 0.19\% degradation. These results highlight that both distribution conditioning via ASH and careful selection of the FP8 format are essential for maintaining stability in TP communication compression.

\begin{table}[b]
\centering
\caption{ASH block size ablation on accuracy and throughput. Throughput is measured in TFLOPS, with speedup relative to the baseline. Bold denotes the best result.}
\label{tab:block_ablation}
\resizebox{\columnwidth}{!}{
\begin{tabular}{lcccc}
\toprule
\textbf{Block Size} & \textbf{Val Loss} $\downarrow$ & \textbf{Test Loss} $\downarrow$ & \textbf{Throughput} $\uparrow$ & \textbf{Speedup} $\uparrow$ \\
\midrule
Baseline & 2.663061 & 2.650814 & 27.1 & 1.00$\times$ \\
ASH (32) & 2.670513 & 2.658282 & 29.5 & 1.09$\times$ \\
ASH (64) & 2.670663 & 2.658583 & 30.2 & 1.11$\times$ \\
ASH (128) & 2.668261 & 2.656122 & 37.9 & 1.40$\times$ \\
\textbf{ASH (256)} & \textbf{2.667557} & \textbf{2.655678} & \textbf{41.2} & \textbf{1.52$\times$} \\
ASH (512) & 2.670552 & 2.658692 & 38.1 & 1.41$\times$ \\
\bottomrule
\end{tabular}
}
\end{table}

\subsubsection{ASH Block Size Sensitivity}

The block size ($B$) of ASH is a critical hyperparameter that governs the trade-off between numerical stability and computational efficiency. Table~\ref{tab:block_ablation} reports the validation loss, test loss, and training performance in a range of block sizes. These results provide several key insights into how the granularity of the transformation affects the statistical conditioning and performance of TP intermediate tensors.

At small block sizes ($B \in \{32, 64\}$), overly fine-grained partitioning introduces non-negligible overhead due to increased kernel invocations. Limited per-block computation leads to poor GPU utilization and suboptimal memory bandwidth efficiency, resulting in only modest throughput gains of $1.09$--$1.11\times$ over the baseline. Moreover, the limited receptive field of the Hadamard transform at this granularity is insufficient to effectively reshape TP intermediate tensor distributions, leading to a slight convergence degradation.

In contrast, moderate block sizes ($B \in {128, 256}$) significantly improve arithmetic intensity and enable more efficient memory coalescing. As reported in Table~\ref{tab:block_ablation}, $B=256$ achieves the highest speedup of $1.52\times$ while preserving sufficient spatial scope to effectively align the reshaped TP intermediate tensor distribution with the FP8 dynamic range. This configuration strikes an optimal balance between computational efficiency and numerical fidelity, while maintaining near-baseline convergence behavior in practice.

However, excessively large blocks ($B=512$) degrade both throughput and numerical accuracy. Relative to the optimal block size of $B=256$, the reduced speedup of $1.41\times$ stems from thread-level workload imbalance and increased shared-memory bank conflicts, which limit effective parallelism and overall execution efficiency. Moreover, the expanded spatial scope of the transform diminishes the effectiveness of adaptive scaling by aggregating TP intermediate tensors with heterogeneous magnitudes into a single scaling region, resulting in a modest but systematic loss of numerical fidelity and slightly impaired convergence.

In summary, a block size of $B=256$ provides the best trade-off between computational efficiency and numerical robustness. Maximize GPU parallelism while ensuring that the reshaped TP intermediate tensor distribution remains well conditioned for FP8 quantization. This balance is crucial for TACO to provide high performance and almost lossless low-bit TP communication.

\begin{figure}[t]
    \centering
    \includegraphics[width=0.95\linewidth]{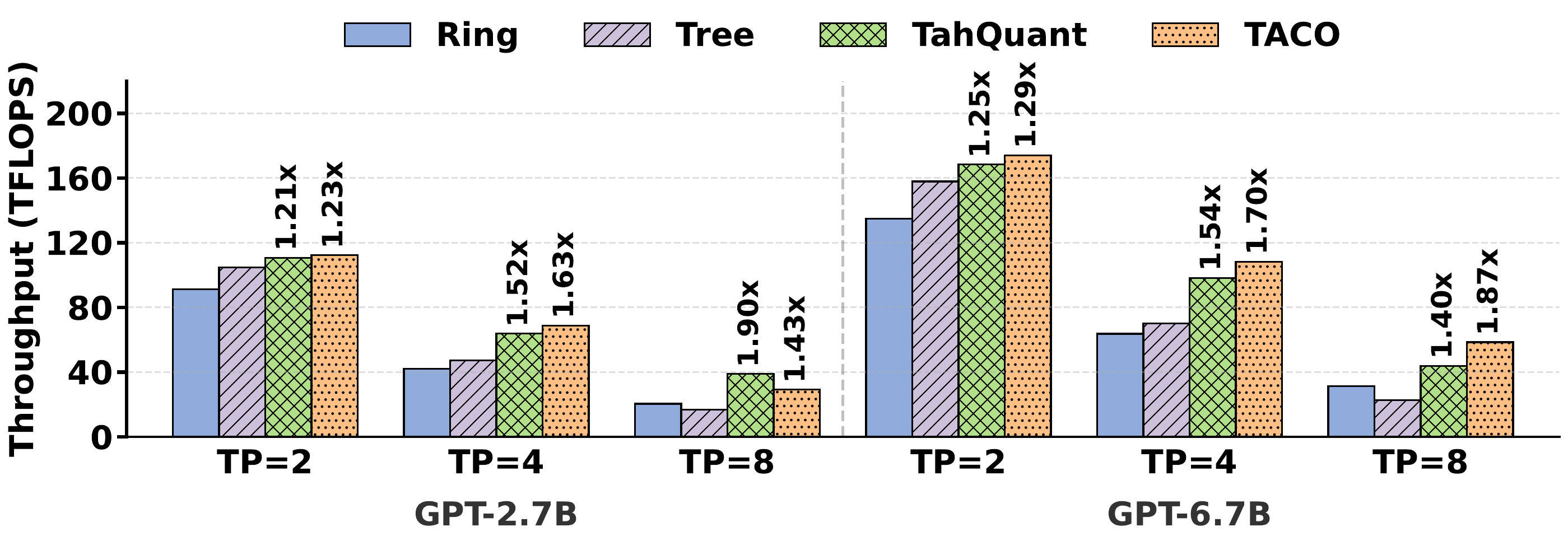}
    \caption{End-to-end training throughput (TFLOPS) under different TP degrees on GPT-2.7B and GPT-6.7B. We compare standard Ring and Tree-based collectives with TahQuant and TACO.}
    \Description{End-to-end training throughput (TFLOPS) under varying tensor parallel degrees for GPT-2.7B and GPT-6.7B. The study compares Ring and Tree-based collective communication with TahQuant and TACO, demonstrating improved scalability and higher throughput of TACO across different model sizes and parallel configurations.}
    \label{fig:tp_throughput}
\end{figure}

\subsection{Evaluation of Performance with TP}
\label{subsec:tp_performance}

We next analyze system-level performance under TP degrees of 2, 4, and 8 on GPT-2.7B and GPT-6.7B. Section~\ref{subsubsec:throughput} to quantify end-to-end throughput and communication scalability under different collective strategies, and Section~\ref{subsubsec:performance_breakdown} to uncover the underlying sources of TACO’s performance gains through a detailed decomposition of computation, communication, and compression overhead.

\subsubsection{End-to-End Throughput Comparison}
\label{subsubsec:throughput}

We systematically evaluate end-to-end training throughput for GPT-2.7B and GPT-6.7B under TP degrees of 2, 4, and 8, comparing Ring AllReduce, tree-based collective communication, TahQuant, and TACO (see Figure~\ref{fig:tp_throughput}. Throughput is measured in TFLOPS, and relative improvements are reported primarily as speedup over the Ring baseline. Ring and tree-based collectives are two widely used communication algorithms in distributed training: Ring AllReduce overlaps communication with computation but incurs latency that scales linearly with the TP degree, whereas tree-based collectives reduce the number of communication steps but often suffer from limited bandwidth utilization and increased synchronization overhead at scale.

Across most configurations, TACO achieves the highest throughput and largest speedups over Ring AllReduce. On GPT-2.7B with TP=2, TACO improves throughput by approximately $1.23\times$ over Ring, slightly surpassing TahQuant ($1.21\times$). As the TP degree increases to 4, communication overhead becomes more pronounced: Ring throughput degrades significantly, whereas TACO maintains a $1.63\times$ speedup, outperforming TahQuant’s $1.52\times$. At TP=8, TahQuant slightly outperforms TACO, achieving a $1.43\times$ speedup versus $1.90\times$ for TACO, reflecting reduced amortization efficiency of TACO’s aggressive kernel optimizations under extreme TP.

In contrast, TACO consistently outperforms all baselines on GPT-6.7B across all TP degrees. At TP=2, it achieves a $1.29\times$ speedup over Ring, surpassing TahQuant’s $1.25\times$. The advantage grows with higher TP degrees: at TP=4, TACO reaches a $1.70\times$ speedup versus $1.54\times$ for TahQuant, and at TP=8, it maintains a $1.87\times$ improvement, compared to $1.40\times$ for TahQuant. These results clearly highlight TACO’s increasing strong efficiency relative to alternatives as TP scales, thanks to its ability to compress TP intermediate tensors and effectively overlap communication with computation.

Overall, throughput decreases as the TP degree increases due to the rapidly growing volume and frequency of TP intermediate tensor communication. TACO consistently mitigates this degradation by compressing TP intermediate tensors into FP8 and fusing compression, decompression, and communication into optimized kernels. By overlapping these operations asynchronously, TACO reduces the effective communication cost along the critical path, improving hardware utilization and delivering substantially better scalability across model sizes and TP configurations.

\begin{figure}[t]
    \centering
    \includegraphics[width=0.95\linewidth]{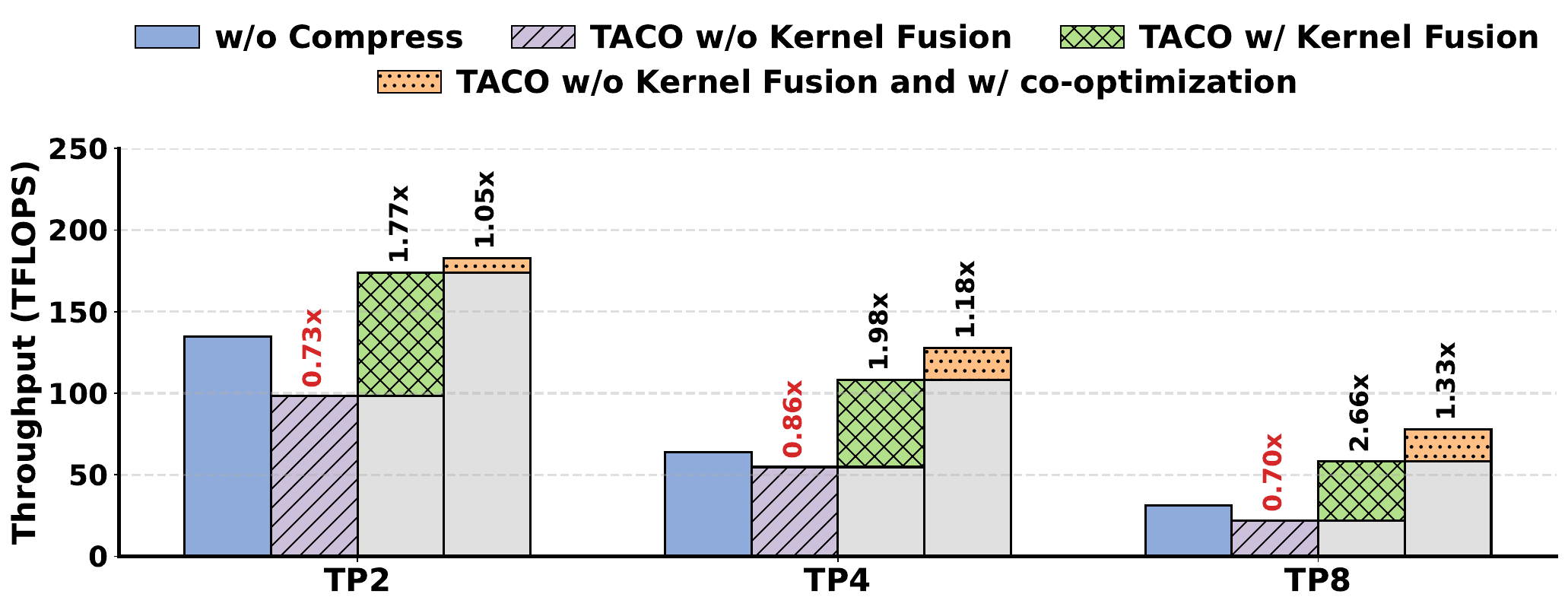}
    \caption{Throughput comparison across different TP settings. The numbers above the bars indicate the speedup ratio relative to the previous baseline.}
    \Description{Throughput comparison across different tensor parallel configurations. The results report both absolute throughput and relative speedup over the baseline, demonstrating consistent performance improvements of the proposed method under various parallel settings.}
    \label{fig:throughput_comparison}
\end{figure}

\subsubsection{Performance Breakdown}
\label{subsubsec:performance_breakdown}

We evaluate TACO’s performance on GPT-6.7B under varying TP degrees, reporting end-to-end throughput in TFLOPS (Figure~\ref{fig:throughput_comparison}). The uncompressed baseline achieves 134.9, 63.7, and 31.3 TFLOPS for TP2, TP4, and TP8, respectively. Applying TACO without kernel fusion slightly reduces throughput due to the local computation overhead of compression, yielding 98.2, 54.8, and 22.0 TFLOPS—corresponding to relative speedups of 0.73$\times$, 0.86$\times$, and 0.70$\times$ compared to the baseline.

Applying kernel fusion dramatically improves performance: TACO with fusion achieves 174.1, 108.3, and 58.5 TFLOPS for TP2, TP4, and TP8, corresponding to speedups of 1.77$\times$, 1.98$\times$, and 2.66$\times$ over TACO without fusion. Further co-optimization on top of kernel fusion provides additional speedups of 1.05$\times$, 1.18$\times$, and 1.33$\times$ for TP2, TP4, and TP8, respectively. These results demonstrate that TACO’s optimizations effectively exploit computation–communication co-design: while FP8 compression introduces minor local computation, kernel fusion and co-optimization maximize throughput, especially as TP communication dominates. Overall, Figure~\ref{fig:throughput_comparison} highlights that TACO consistently delivers substantial performance gains over the  baseline, with the benefits of combined optimizations growing as TP increases.

\subsection{3D Parallel Training Evaluation}
\label{sec:3d_system_eval}

\subsubsection{Training Accuracy}

\begin{figure}[t]
    \centering
    \includegraphics[width=1.0\linewidth]{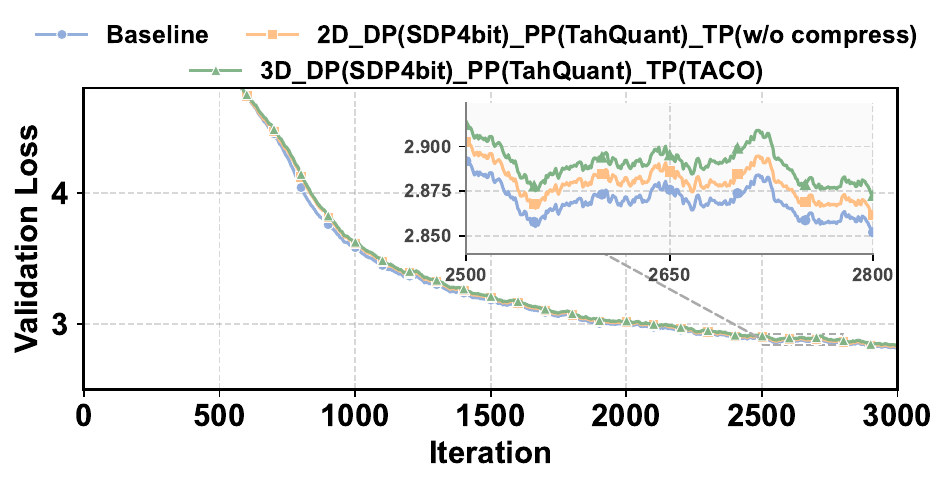}
    \caption{Validation loss of GPT-6.7B under full 3D parallelism.}
    \Description{Validation loss curves of GPT-6.7B under full 3D parallel training. The results demonstrate stable convergence of the proposed approach, indicating strong numerical robustness and training consistency at large scale.}
    \label{fig:gpt_loss}
\end{figure}

We evaluate TACO’s accuracy under full 3D parallelism on GPT-6.7B. To isolate the impact of TP compression, we consider three settings applied consistently to both models: (1) a baseline without compression, (2) 2D parallelism, where DP and PP communications are quantized using SDP4bit and TahQuant respectively while TP remains uncompressed, and (3) 3D parallelism, where TACO additionally use in TP alongside SDP4bit and TahQuant.
Figure~\ref{fig:gpt_loss} presents the validation loss curves of GPT-6.7B under these settings. The baseline model achieves a final loss of 2.663061, while the 2D-parallel configuration slightly degrades to 2.676003 due to quantization noise in DP and PP communications. In contrast, TACO under full 3D parallelism closely tracks the baseline throughout training, reaching a final loss of 2.679906, corresponding to only a 0.14\% loss increase over the 2D setting, demonstrating that adding TP compression does not introduce additional optimization instability.
Across the entire training trajectory, TACO exhibits stable convergence behavior comparable to full-precision training, indicating that the proposed method effectively preserves gradient fidelity even under aggressive communication compression across all parallel dimensions.
These results highlight that TACO enables end-to-end compression of DP, PP, and TP communications in large-scale training while maintaining near-lossless optimization quality, demonstrating strong robustness and scalability in full 3D parallel training settings, and consistently delivering reliable performance across different model scales and training configurations.

\subsubsection{End-to-End Throughput}
\label{subsec:3d_performance}

Finally, we present an end-to-end training throughput evaluation of TACO on GPT models. As shown in Table~\ref{tab:3d_tflops}, compared with the uncompressed baseline, TACO achieves consistent and significant performance improvements across different model sizes.
For GPT models, the speedup reaches up to \textbf{1.53$\times$}. Notably, the improvements from 2D to full 3D parallelism are substantially amplified once TP communication compression is enabled, indicating that tensor parallel communication is a major bottleneck in large-scale distributed training.

These results demonstrate that, under tightly synchronized 3D parallel training, reducing TP communication volume is critical for achieving high system efficiency. TACO’s optimized compression operators ensure that the additional quantization overhead is fully amortized by the reduction in communication cost, resulting in a net throughput gain. Moreover, the consistent speedups across GPT model scales suggest that TACO scales robustly with increasing model size and communication intensity.
Training remains numerically stable throughout optimization. As shown in Figure~\ref{fig:gpt_loss}, TACO preserves a validation loss trajectory that closely matches the full-precision baseline, confirming that the system-level gains are achieved without sacrificing convergence or accuracy.

\section{Discussion}

\textbf{Design objective.}
TACO is not designed to maximize raw communication speed, but to enable \textbf{near-lossless} compression of TP intermediate tensors while preserving convergence in large-scale training. This is motivated by the observation that existing methods often introduce optimization instability under high-frequency synchronization or require delicate tuning of error compensation and scaling strategies, limiting their robustness.
To achieve this, TACO adopts FP8 as a practical precision point, striking a balance between compression efficiency and numerical fidelity, and enabling stable training under aggressive communication reduction.

\textbf{Generality across hardware.}
Although TACO is implemented with FP8 as the primary precision format, its design is not dependent on FP8-specific hardware support. Instead, FP8 serves as a target precision level that defines the compression semantics of TP communication. On platforms without native FP8 support, TACO degrades gracefully to an INT8-based implementation, where quantization is performed using our ASH together with DS.
In this configuration, intermediate tensors are still compressed to low-bit integer representations during communication, while scaling and reconstruction are handled in a software-assisted manner. This design preserves the same communication semantics as FP8-based execution, while trading off additional lightweight arithmetic overhead for broader hardware compatibility. As a result, TACO maintains its communication reduction benefits across heterogeneous accelerators without requiring specialized FP8 support.

\begin{table}[t]
\centering
\caption{End-to-end throughput (TFLOPS) under 3D parallelism on GPT models (TP=4, PP=2, DP=2).}
\label{tab:3d_tflops}
\small 
\begin{tabular}{lcccc}
\toprule
\textbf{Models} & \textbf{Size} & \textbf{Baseline} & \textbf{2D (w/o TACO)} & \textbf{3D (w/ TACO)} \\
\midrule
\multirow{3}{*}{GPT} 
 & 2.7B  & 39.9  & 40.3 (1.01$\times$)  & 59.7 (\textbf{1.50$\times$})  \\
 & 6.7B  & 61.1  & 62.3 (1.02$\times$)  & 93.3 (\textbf{1.53$\times$})  \\
 & 13B   & 73.8  & 75.2 (1.02$\times$)  & 111.7 (\textbf{1.51$\times$}) \\
\bottomrule
\end{tabular}
\end{table}

\section{Conclusion and Future Work}
\label{sec:conclusion}

In this paper, we presented \textbf{TACO}, an efficient framework for accelerating distributed training by optimizing the communication of TP intermediate tensors. By leveraging FP8-based compression and system-level kernel fusion, TACO effectively alleviates the communication bottlenecks inherent in high-degree TP. Our systematic evaluations across multiple model scales, including GPT and Qwen, demonstrate that TACO consistently achieves superior throughput, with speedups of up to 1.87$\times$. Moreover, experiments under full 3D parallelism confirm that TACO delivers robust, architecture-agnostic performance gains while maintaining convergence nearly identical to uncompressed baselines.
Future work will focus on extending \textbf{TACO}'s compression strategies to encompass gradient and optimizer state communication within 3D-parallel training stacks. Additionally, we aim to investigate adaptive quantization schemes that dynamically adjust the precision of TP intermediate tensors according to layer-wise sensitivity, further enhancing hardware utilization and efficiency in exascale distributed training systems.

\begin{acks}
This work was supported by the National Key Research and Development Program of China (Grant No. 2025YFB3003702), the Innovation Funding of ICT, CAS (Grant No. E461050), and the National Natural Science Foundation of China (Grant Nos. 62032023 and T2125013). The AI-driven experiments, simulations, and model training were conducted on the robotic AI-Scientist platform at the Chinese Academy of Sciences.
\end{acks}

\newpage
\balance
\bibliographystyle{ACM-Reference-Format}
\bibliography{reference}

\end{document}